\documentclass[12pt]{article}
\usepackage{epsfig}
\newcommand{\pfrac}[2]{\left(\frac{#1}{#2}\right)}

\begin{document}
\begin{flushright}
MZ-TH/06-09\\
nlin.CD/0609052\\
May 2006\\
\end{flushright}

\begin{center}
{\Large\bf Scaling properties of invariant densities\\[0.3truecm]
  of coupled Chebyshev maps} \\[1truecm]
{\large Stefan Groote$^{1,2}$ and Christian Beck$^3$}\\[.7cm]

$^1$ Institut f\"ur Physik der Johannes-Gutenberg-Universit\"at,\\
  Staudinger Weg 7, 55099 Mainz, Germany\\[.5truecm]
$^2$ F\"u\"usika-Keemiateaduskond, Tartu \"Ulikool,
  T\"ahe 4, 51010 Tartu, Estonia\\[.5truecm]
  $^3$ School of Mathematical Sciences, Queen Mary,\\
  University of London, Mile End Road, London E1 4NS, UK\\[.5truecm]
\vspace{1truecm}
\end{center}

\begin{abstract}
We study 1-dimensional coupled map lattices consisting of diffusively coupled
Cheby\-shev maps of $N$-th order. For small coupling constants $a$ we
determine the invariant 1-point and 2-point densities of these nonhyperbolic
systems in a perturbative way. For arbitrarily small couplings $a>0$ the
densities exhibit a selfsimilar cascade of patterns, which we analyse in
detail. We prove that there are log-periodic oscillations of the density both
in phase space as well as in parameter space. We show that expectations of
arbitrary observables scale with $\sqrt{a}$ in the low-coupling limit,
contrasting the case of hyperbolic maps where one has scaling with $a$.
Moreover we prove that there are log-periodic oscillations of period
$\log N^2$ modulating the $\sqrt{a}$-dependence of the expectation value of
any given observable.
\end{abstract}

\newpage

\section{Introduction}
Coupled map lattices (CMLs) as introduced by Kaneko and Kapral
\cite{kaneko,kapral} are interesting examples of spatially extended dynamical
systems that exhibit a rich structure of complex dynamical phenomena
\cite{kanekobook,Beck:2002,Bunimovich:1988,amritkar,carretero}. Both space and
time are discrete for these types of dynamical systems, but the local field
variable at each lattice site takes on continuous values. CMLs have a variety
of applications: They have been studied e.g.\ as models for hydrodynamical
flows, turbulence, chemical reactions, biological systems, and quantum field
theories (see e.g.\ reviews in~\cite{kanekobook,Beck:2002}). Of particular
interest are CMLs that exhibit spatio-temporal chaotic behaviour. The analysis
of these types of CMLs is often restricted to numerical investigations and
only a few analytical results are known. A notable exception is the case of
CMLs consisting locally of hyperbolic maps for small coupling $a$. Here a
variety of analytical results is known
\cite{bunimovich,baladi,Bricmont:1996,Jarvenpaa:2001} that guarantee the
existence of a smooth invariant density and ergodic behaviour.

The situation is much more complicated for nonhyperbolic local maps which
correspond to the generic case of physical interest \cite{daido,Lemaitre:1998,%
Lemaitre:1999,ding,ruffo,mackey,beck}. In the one-dimensional case,
non-hyperbolicity simply means that the absolute value of the slope of the
local map is allowed to be smaller than 1 in some regions. For chaotic CMLs
consisting of nonhyperbolic maps much less is known analytically, and standard
techniques from ergodic theory are usually not applicable. From the
applications point of view, the nonhyperbolic case is certainly the most
interesting one. For example, it has been shown that nonhyperbolic CMLs
exhibiting spatio-temporal chaos naturally arise from stochastically quantised
scalar field theories in a suitable limit where the strength of the underlying
self-interacting potentials goes to infinity~\cite{Beck:2002,Beck:1995}. CMLs
obtained in this way can serve as useful models for vacuum fluctuations and
dark energy in the universe~\cite{dark}. Other applications include chemical
kinetics as described by discretised reaction diffusion
dynamics~\cite{kanekobook}.

In this paper, we study weakly coupled $N$-th order Chebyshev maps
\cite{beck,Beck:1995,dark,dettmann1,dettmann2,groote}. We consider CMLs with
diffusive coupling, i.e.\ there is a coupling between neighboured maps that
resembles a discrete version of the Laplacian on the lattice. This form of the
coupling is the most important one for the quantum field theoretical
applications \cite{Beck:2002}. In the uncoupled case it is well known that the
Chebyshev maps are conjugated to a Bernoulli shift of $N$ symbols (see,
e.g.\ \cite{Beck:1993}). In the coupled case, this conjugacy is destroyed and
the conventional treatments for hyperbolic maps does not apply, since the
Chebyshev maps have $N-1$ critical points where the slope vanishes, thus
corresponding to a nonhyperbolic situation.

We will derive explicit formulas for the invariant densities describing the
ergodic behaviour of weakly coupled Chebyshev systems, by applying a
perturbative treatment. Our analytical techniques will yield explicit
perturbative expressions for the invariant 1-point and 2-point density for
small couplings $a$. We will prove a variety of interesting results concerning
the scaling behaviour of these nonhyperbolic systems. Our main results are
that the 1-point density exhibits log-periodic oscillations of period
$\log N^2$ near the edges of the interval. Expectations of typical observables
scale with $\sqrt{a}$ (rather than with $a$ as for hyperbolic coupled maps).
We also show that there are log-periodic modulations of arbitrary expectation
values as a function of the coupling parameter $a$. Although our results are
rigorously derived for the explicit example of Chebyshev maps only, we
expect that similar results hold for other nonhyperbolic systems as well,
since our techniques are quite general and can be applied to various other
systems in a similar way.

This paper is organized as follows: In section 2 we introduce the relevant
class of coupled map lattices and provide some numerical evidence for the
observed scaling phenomena, choosing as a main example the Ulam map (the $N=2$
Chebyshev map). Our perturbative theory of the invariant density of the
coupled system is presented in section 3. Based on these results we explain
details of the selfsimilar patterns in section 4. In section 5 we extend our
results to the third order Chebyshev map, where, due to an additional
discrete symmetry, the behaviour is somewhat different as compared to $N=2$.
Finally, in section 6 we collect our results to finally prove that there are
log-periodic oscillations modulating the scaling behaviour of expectation
values of arbitrary observables if the coupling parameter $a$ is changed.

\section{Observed scaling phenomena}
We consider one-dimensional coupled map lattices of the form
\begin{equation}\label{evolution}
\phi_{n+1}^i=(1-a)T_N(\phi_n^i)+s\frac{a}2\left(T_N^b(\phi_n^{i+1})
  +T_N^b(\phi_n^{i-1})\right).
\end{equation}
Here $T_N(\phi)$ denotes the $N$-th order Chebyshev polynomial and
$\phi_n^i$ is a local iterate at lattice site $i$ at time $n$. Our main
interest is concentrated on the cases $N=2$ where $T_2(\phi)=2\phi^2-1$ and
$N=3$ where $T_3(\phi)=4\phi^3-3\phi$. The sign $s=\pm 1$ distinguishes
between the `diffusive coupling' ($s=+1$) and the `anti-diffusive coupling'
($s=-1$). The index $b$ finally can take the values $b=0,1$ where $b=1$
corresponds to forward coupling ($T_N^1(\phi):=T_N(\phi)$) whereas $b=0$
stands for backward coupling ($T_N^0(\phi):=\phi$). The discrete chaotic noise
field variables $\phi_n^i$ take on continuous values on the interval $[-1,1]$.
The initial values $\phi_0^i$ are randomly chosen in this interval with a
smooth density, usually chosen to be the invariant density of the uncoupled
system.

\subsection{Scaling in parameter space}
The variable $a$ in Eq.~(\ref{evolution}) is a coupling parameter which takes
values in the interval $[0,1]$. It determines the strength of the Laplacian
coupling in Eq.~(\ref{evolution}) (in the particle physics applications
described in \cite{Beck:2002} $a^{-1}$ can be regarded as a kind of metric for
a one-dimensional `chaotic string'). For $a=0$ we end up with the chaotic
behaviour $\phi_{n+1}^i=T_N(\phi_n^i)$ as generated by uncoupled Chebyshev
maps. If a small coupling $a>0$ is switched on, one observes the formation of
nontrivial density patterns with scaling behaviour with respect to $\sqrt{a}$.
Consider an arbitrary test function (observable) $V(\phi)$ of the local
iterates, and let $\langle\ldots\rangle_a$ denote the expectation values
formed by long-term iteration and averaging over all lattice sites. For small
$a$ one observes numerically
\begin{equation}
\langle V(\phi)\rangle_a-\langle V(\phi)\rangle_0=f^{(N)}(\ln a)a^{1/2}
\end{equation}
where $f^{(N)}(\ln a)$ is a periodic function of $\ln a$ with period $\ln N^2$.
The function $f^{(N)}=f^{(N)}[V]$ is a functional of $V$. A double logarithmic
plot of $|\langle V(\phi)\rangle_a-\langle V(\phi)\rangle_0|$ versus $a$ hence
shows a straight line that is modulated by oscillating behaviour. Examples are
shown in Fig.~1.
\begin{figure}\begin{center}
\epsfig{file=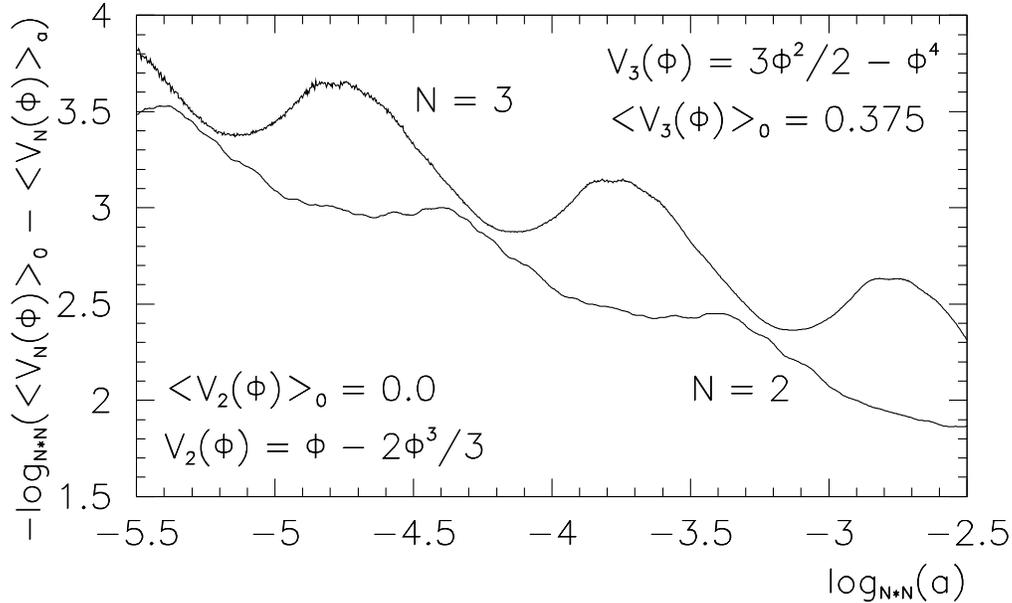, scale=0.8}
\caption{Scaling in parameter space for the test functions
  $V(\phi)=V_2(\phi):=\phi-\frac23\phi^3$ ($N=2$) and
  $V(\phi)=V_3(\phi):=\frac32\phi^2-\phi^4$ ($N=3$). To calculate the
  expectations, the coupled maps were iterated $10^5$ times for random initial
  conditions. The lattice size was $10^3$.}
\end{center}\end{figure}
We will rigorously prove the observed log-periodic scaling behaviour with
respect to $a$ in section 6 of this paper.

For small $a$ there is no difference between expectations calculated with the
forward or backward coupling form of (\ref{evolution}), the value of $b$
matters only for large values of the coupling.

\subsection{Scaling in the phase space}
There is also scaling behaviour in the phase space of the coupled system.
First, let us consider the uncoupled case. The invariant (one-point
propability) density corresponding to $a=0$ is known to be
\begin{equation}\label{rho0}
\rho_0(\phi)=\frac1{\pi\sqrt{1-\phi^2}}
\end{equation}
(see e.g.\ \cite{Beck:1993} for a derivation). This density is universal in
the sense that it does not depend on the index $N$ of the Chebyshev
polynomial for $N\geq 2$.

The 1-point invariant density at each lattice site will of course change if we
consider values $a\ne 0$. It then depends not only on $a$ but also on the
index $N$. In other words, the $N$-degeneracy is broken for finite coupling.
In principle, the invariant densities of coupled map lattices can be
understood by finding fixed points of the Perron-Frobenius operator of these
very high-dimensional dynamical
systems~\cite{Bricmont:1996,Baladi:2000,Keller:1992,Lemaitre:1997,Chate:1997}.
The 1-point density can then be obtained as a marginal distribution, by
integrating out all lattice degrees of freedom up to one. In practice, this
program is rather complicated for our type of systems and we will apply a more
pedestrian but feasable method in the following sections.

\begin{figure}\begin{center}
\epsfig{figure=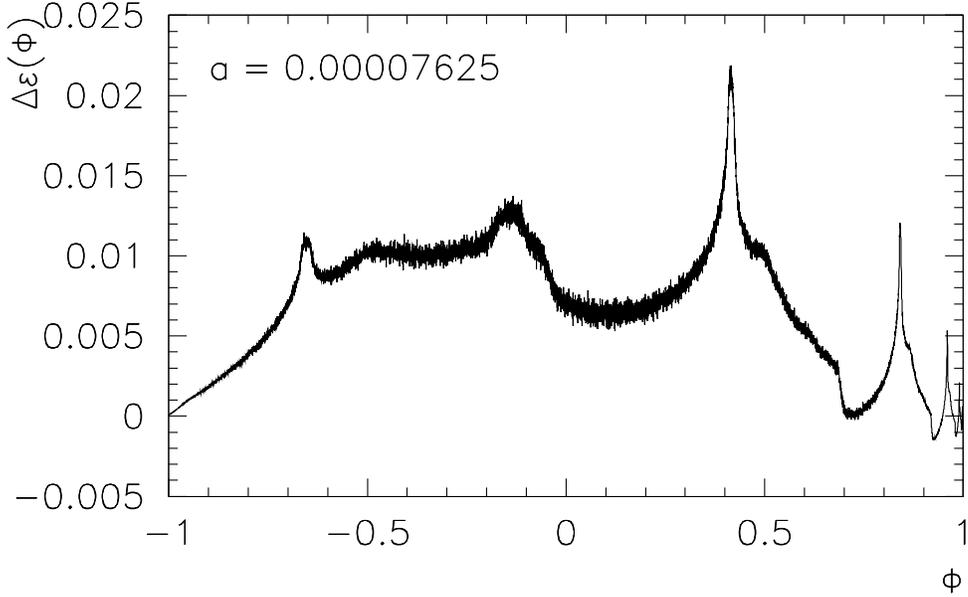, scale=0.8}
\caption{\label{ener23}Energy difference $\Delta\epsilon(\phi)$
for the
  2B-string at $a=0.00007625$}
\end{center}\end{figure}

If the coupling $a$ is slightly increased starting from $a=0$, the invariant
1-point density $\rho_a(\phi)$ starts to deviate from $\rho_0(\phi)$. There
are particularly strong deviations at the boundaries $\phi=\pm 1$ of the
interval. While for odd $N$ the evolution equation~(\ref{evolution}) is
invariant under the replacement $\phi\to-\phi$ and hence the invariant density
is symmetric, this is not the case for even values of $N$. In both cases we
are interested in the region close to $\phi=\pm 1$ and our goal is to
understand the deviation of $\rho_a(\phi)$ from $\rho_0(\phi)$.

For our purpose it is useful to define an `effective energy' $\epsilon_a$ as
follows:
\begin{equation}
\epsilon_a(\phi):=\frac12-\frac1{2\pi^2\rho_a(\phi)^2}.
\end{equation}
This function is a useful tool if one wants to apply generalized statistical
mechanics techniques to coupled Chebyshev maps, see~\cite{Beck:2002} for more
details. For $a=0$ the above effective energy is just given by the `kinetic
energy' expression $\frac12\phi^2$ of a free particle, and the invariant
density coincides with that of a generalized canonical ensemble in Tsallis'
formulation of nonextensive statistical mechanics~\cite{tsallis}, the entropic
index $q$ being given by $q=3$. For $a\neq 0$ the effective energy becomes
more complicated and it includes the effects of interactions with neighboured
lattice sites. The quantity of interest is the energy difference
$\Delta\epsilon(\phi)=\epsilon_a(\phi)-\epsilon_0(\phi)$.

In~\cite{Beck:2002} a general classification scheme of the relevant
Chebyshev CMLs was introduced. To be specific, let us here consider the
coupled Chebyshev map system with $N=2$, $s=+1$ and $b=0$ (called 2B-string
in~\cite{Beck:2002}). For finite $a$ we obtain an (asymmetric) energy
difference $\Delta\epsilon(\phi)$ which when approaching the boundary
$\phi=+1$ shows selfsimilar behaviour, i.e.\ a repetition of nearly the same
pattern if a suitable rescaling is done (see Fig.~\ref{ener23}).

\begin{figure}\begin{center}
\epsfig{figure=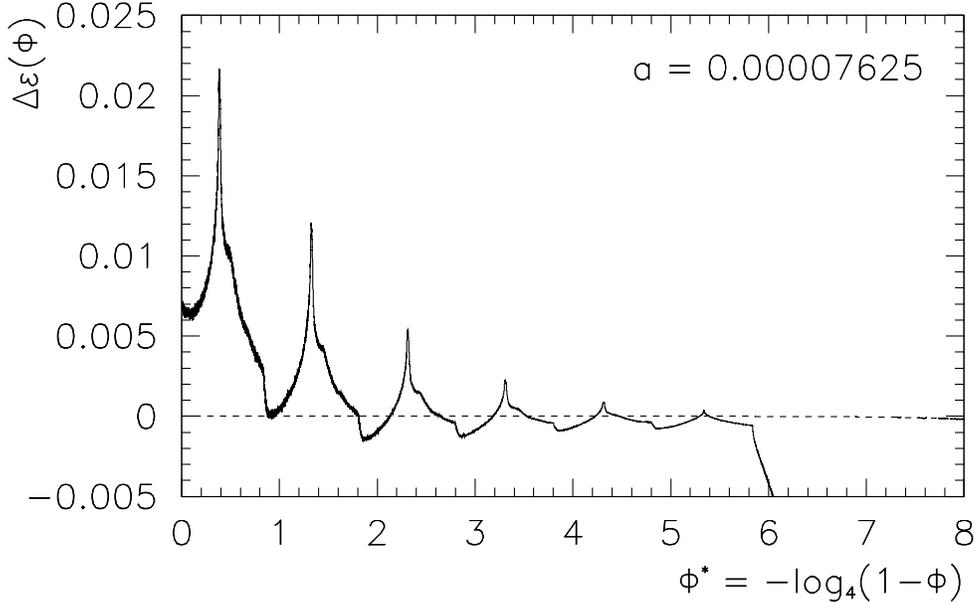, scale=0.8}
\caption{\label{ener23l}Energy difference $\Delta\epsilon(\phi)$ for the
  2B-string for $a=0.00007625$ (logarithmic plot)}
\end{center}\end{figure}

The scaling behaviour close to $\phi=+1$ can be analysed in more detail by
looking at a logarithmic plot of $\Delta\epsilon(\phi)$ versus
$\phi^*=-\log_4(1-\phi)$, as shown in Fig.~\ref{ener23l}. In order to deal
with a logarithmic histogram plot the data are adjusted in the usual way by
the factor $1/(1-\phi)\ln 4$, as obtained from transformation of random
variables. We see that the pattern in this plot is repeated with a period
$\Delta\phi^*=1$, though there is a sharp breakdown of periodicity at the
value $\phi^*=a^*=-\log_4(4a)$. We will prove this behaviour later.

\begin{figure}\begin{center}
\epsfig{figure=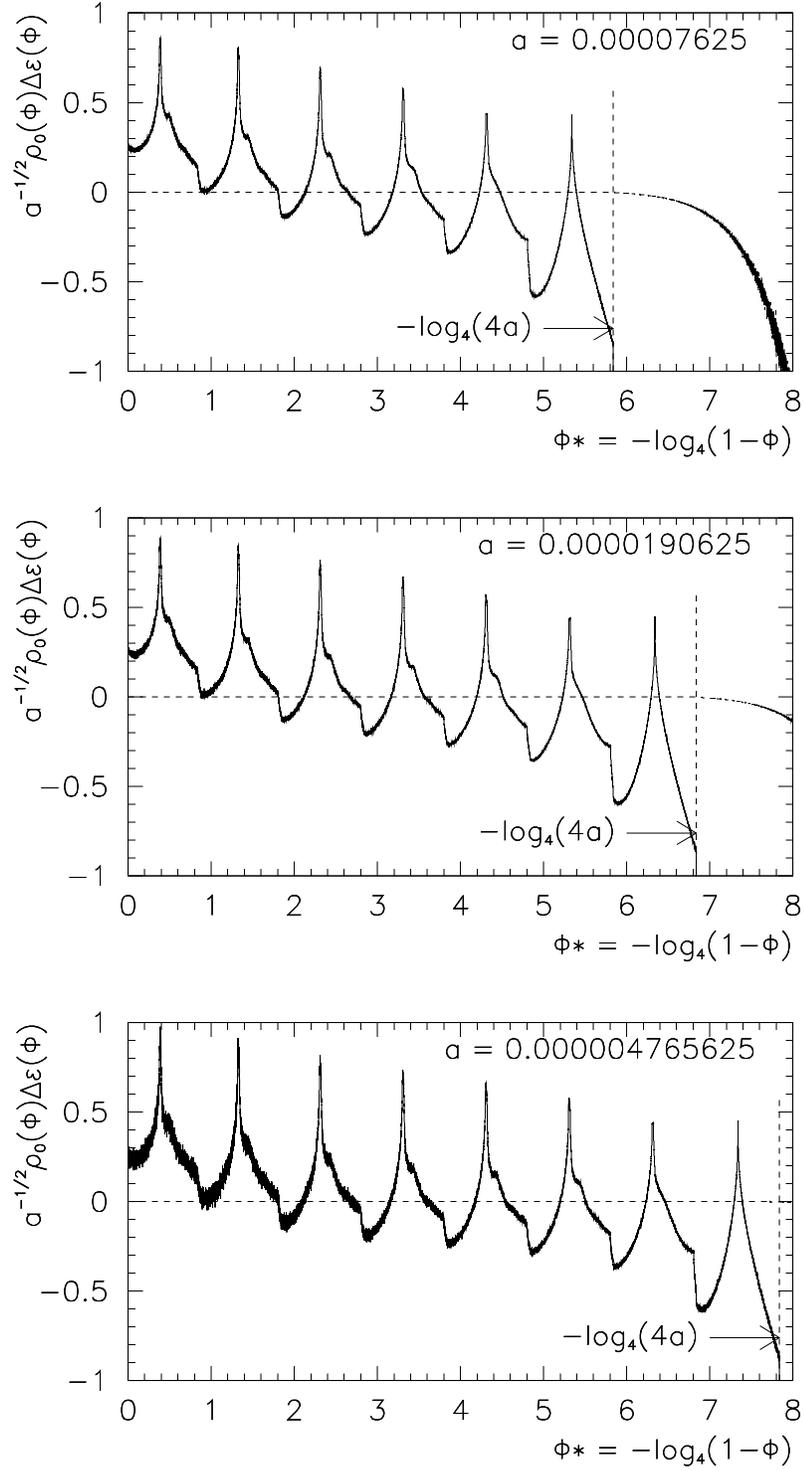, scale=0.75}
\caption{\label{ener2345}Adjusted energy difference
  $a^{-1/2}\rho_0(\phi)\Delta\epsilon(\phi)$ for the 2B-string at
  $a=0.00007625$, $a=0.0000190625$, and $a=0.000004765625$ (from top to
  bottom). Indicated is the breakdown point of the pattern at
  $\phi^*=a^*=-\log_4(4a)$ (vertical line) as well as the pattern amplitude
  multiplied by a factor of 0.001 (horizontal line) in order to have a
  suitable scale for the breakdown region.}
\end{center}\end{figure}

\subsection{Connection between parameter and phase space scaling}
Let us now look at how the energy difference changes with the coupling
parameter $a$. If the coupling is decreased by a factor of $4$ one observes
that the pattern in the logarithmic plot is moved to the right by
$\Delta\phi^*=1$ while a new pattern element is created to the left. The
picture is like a wave generator located at $\phi^*=0$ that was started at
some `time' in the past and is sending out waves with a speed $\Delta\phi^*=1$.
In this analogy `time' corresponds to the negative logarithm of the coupling
parameter. Since in Fig.~\ref{ener23l} the pattern amplitude becomes smaller
and smaller towards higher values of $\phi^*$ we can try to compensate this
effect by multiplying $\Delta\epsilon(\phi)$ by an appropriate simple function
of $\phi^*$. It turns out that the multiplication by $\rho_0(\phi)$
compensates the decrease of the pattern for a fixed value of the coupling $a$.
However, the pattern as a whole turns out to decrease by a factor of $2$ if we
divide the coupling by a factor of $4$. Therefore, as a last adjustment, we
multiply with the inverse square root $1/\sqrt a$ of the coupling. In a
sequence of steps $\Delta a^*=1$ for the `time variable' $a^*=-\log_4(4a)$ we
show in Fig.~\ref{ener2345} the progressive motion of the adjusted pattern,
indicating the breakdown point at $\phi^*=a^*$.

\begin{figure}\begin{center}
\epsfig{figure=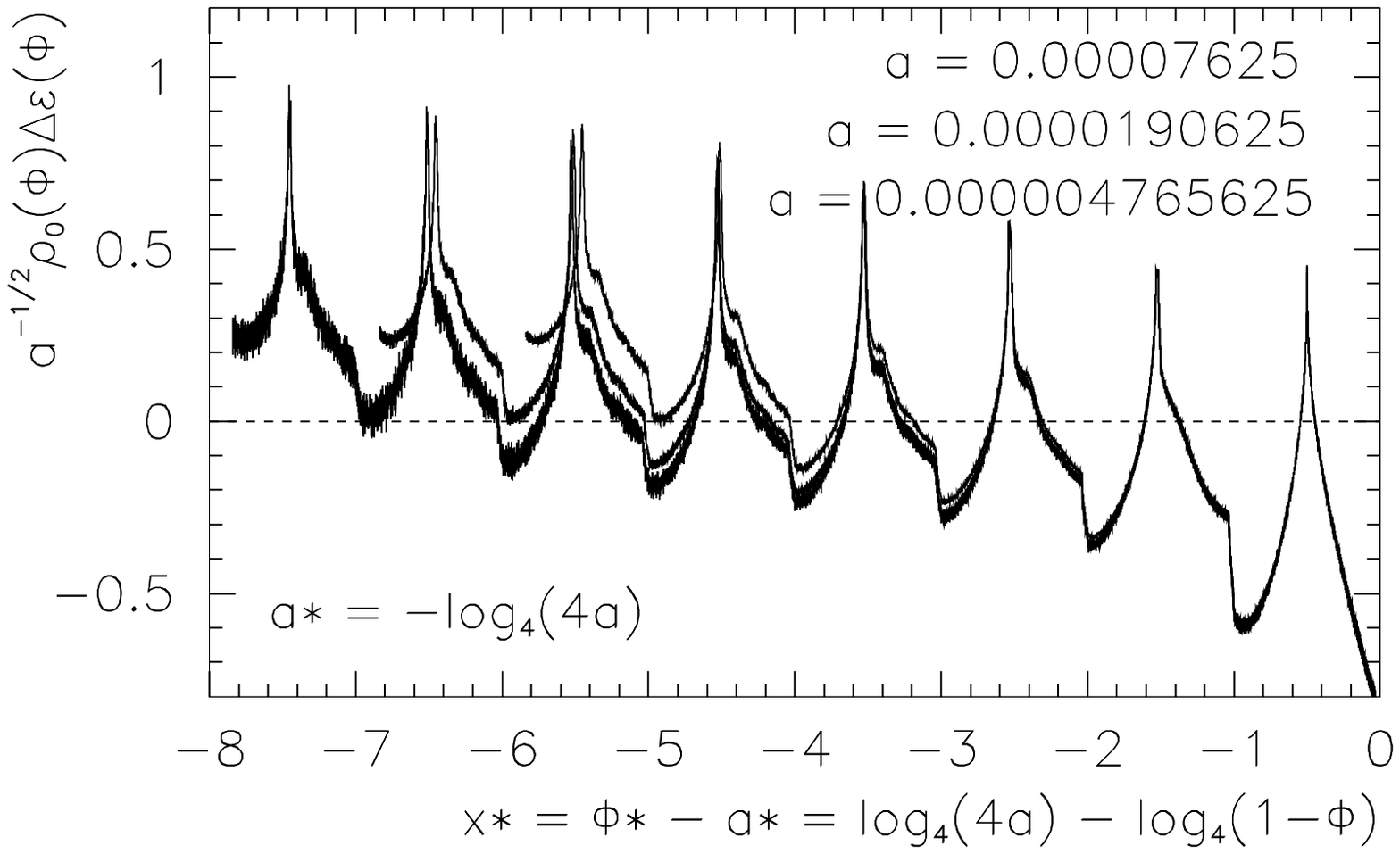, scale=0.75}
\caption{\label{ener2s}Adjusted energy difference
  $a^{-1/2}\rho_0(\phi)\Delta\epsilon(\phi)$ for the 2B-string at
  $a=0.00007625$, $a=0.0000190625$, and $a=0.000004765625$ (from top to
  bottom) as a function of $x^*=\phi^*-a^*$.}
\end{center}\end{figure}

We can compensate the shift of the pattern to the right by subtracting
$a^*=-\log_4(4a)$ from $\phi^*=-\log_4(1-\phi)$ obtaining the new variable
$x^*=\phi^*-a^*$. Note that $x^*$ is negative beyond the breakdown region. In
writing subsequent curves into the same diagram we obtain an entire set of
curves shown in Fig.~\ref{ener2s}.
While the curves diverge for large absolute values of $x^*$, the surprising
fact is that they exactly coincide close to the breakdown at $x^*=0$. This
means that there is invariance under a suitable renormalization transformation:
$a^{-1/2}\rho_0(\phi)\Delta\epsilon(\phi)$ is a function of $x^*=-\log_4(x)$
only if the absolute value of $x^*$ is small. The shape of the first peak
(i.e.\ for $x^*\in[-1,0]$) seems to be very simple, while the following peaks
increase in complexity.

\begin{figure}\begin{center}
\epsfig{figure=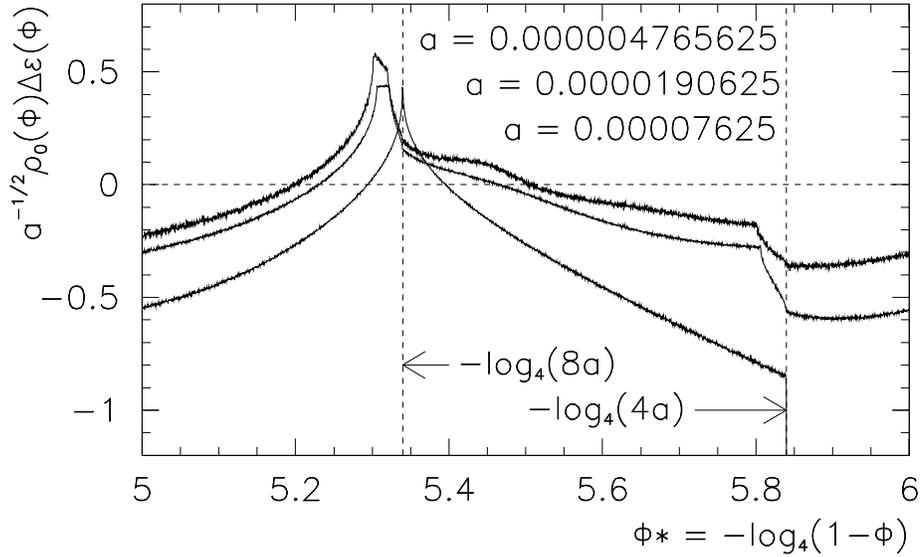, scale=0.75}
\caption{\label{ener2n}Comparison of the first three peaks, as
obtained for a fixed interval of $\phi^*$ with
   $a=0.00007625$, $a=0.0000190625$, and $a=0.000004765625$.
  The indicated logarithms in the
  diagram are related to the value $a=0.00007625$.}
\end{center}\end{figure}

As a last diagram illustrating this behaviour, in Fig.~\ref{ener2n} we show
the development of the density pattern for a given fixed interval of $\phi^*$
for the already used sequence of values of $a$. Note the slight shift of the
maxima to the left and the formation of plateaus. Various details of this
figure will become clearer by our theoretical treatment in section 4.

\section{Perturbative calculation of the invariant density}
\begin{figure}\begin{center}
\epsfig{figure=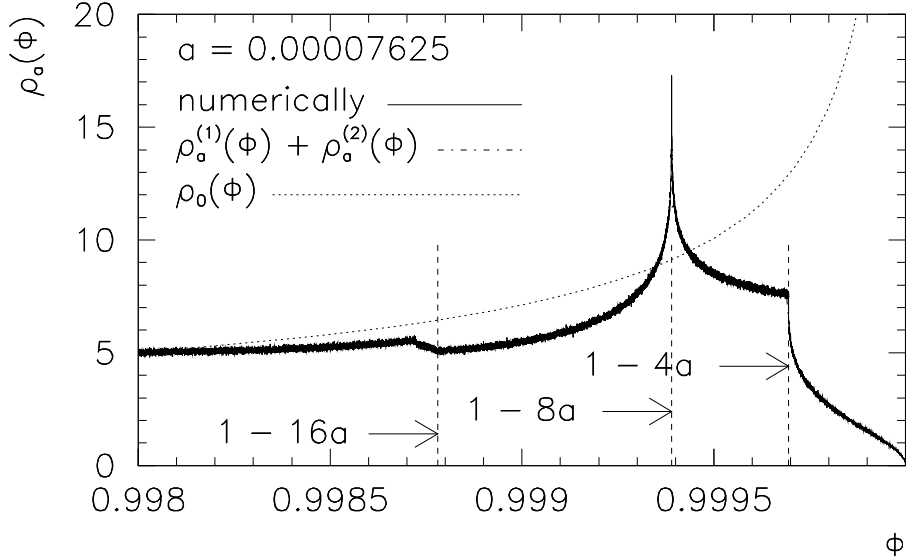, scale=0.75}
\caption{\label{rho23d}Invariant density $\rho_a(\phi)$ close to $\phi=+1$
  for the 2B-string for a=0.00007625, as compared to the invariant
  density $\rho_0(\phi)$ for vanishing coupling (dotted curve).}
\end{center}\end{figure}

Let us now proceed to an analytic (perturbative) calculation of the invariant
1-point density of the 2B-dynamics
\begin{equation}\label{evolution2B}
\phi_{n+1}^i=(1-a)T_2(\phi_n^i)+\frac{a}2(\phi_n^{i+1}+\phi_n^{i-1}).
\end{equation}
We are particularly interested in the behaviour near the boundaries
$\phi=\pm 1$. The numerically determined density near $\phi=+1$ is shown in
Fig.~\ref{rho23d}.

\begin{figure}\begin{center}
\epsfig{figure=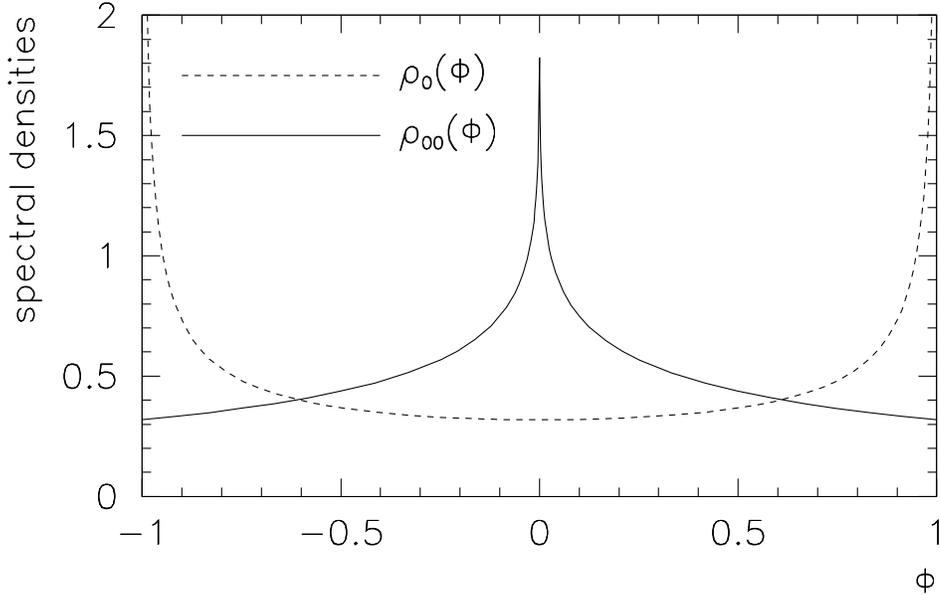, scale=0.8}
\caption{\label{rho000}Invariant densities $\rho_{00}(\phi)$ (solid line) and
  $\rho_0(\phi)$ (dashed line)}
\end{center}\end{figure}

As a first approximation, we essentially regard the system (\ref{evolution2B})
as a perturbed one-dimensional system, the perturbation of strength $a$ given
by the arithmetic mean of nearest neighbours
$(\phi^{i+1}_n+\phi^{i-1}_n)/2=:(\phi_++\phi_-)/2=:\tilde\phi$. As a first
approximation we may also regard $\phi_+$ and $\phi_-$ to be nearly
independent for small $a$, provided we are not too close to the edges of the
interval. This assumption is necessary as a starting point for our
perturbative approach, corrections to this assumption are expected to appear
at higher orders of the perturbation series. The probability density
$\rho_{aa}$ of $\tilde\phi$ is obtained by the convolution
\begin{equation}\label{rhoaaint}
\rho_{aa}(\tilde\phi)=2\int\rho_a(2\tilde\phi-\phi_-)\rho_a(\phi_-)d\phi_-,
\end{equation}
the factor 2 arising from the fact that $d\phi_+d\phi_-=2d\tilde{\phi}d\phi_-$.
If we replace the densities $\rho_a$ occurring in Eq.~(\ref{rhoaaint}) by
$\rho_0$, we can calculate the convolution analytically to obtain
\begin{equation}\label{rho00}
\rho_{00}(\tilde\phi)=\int\frac{2d\phi_-}{\pi^2\sqrt{1-(2\tilde\phi-\phi_-)^2}
  \sqrt{1-\phi_-^2}}=\frac2{\pi^2}K\left(\sqrt{1-\tilde\phi^2}\right)
  \theta(1-\tilde\phi^2),
\end{equation}
where $K(x)$ is the complete elliptic integral of the first kind and
$\theta(x)$ is the step function. The functions $\rho_0(\phi)$ and
$\rho_{00}(\phi)$ are shown in Fig.~\ref{rho000}. If we calculate to leading
order, we can replace $\rho_a(\phi)$ and $\rho_{aa}(\phi)$ by $\rho_0(\phi)$
and $\rho_{00}(\phi)$, respectively. For our perturbative approach we always
assume that the difference between the perturbed and unperturbed density is
small, except at the edges of the interval.

\begin{figure}\begin{center}
\epsfig{figure=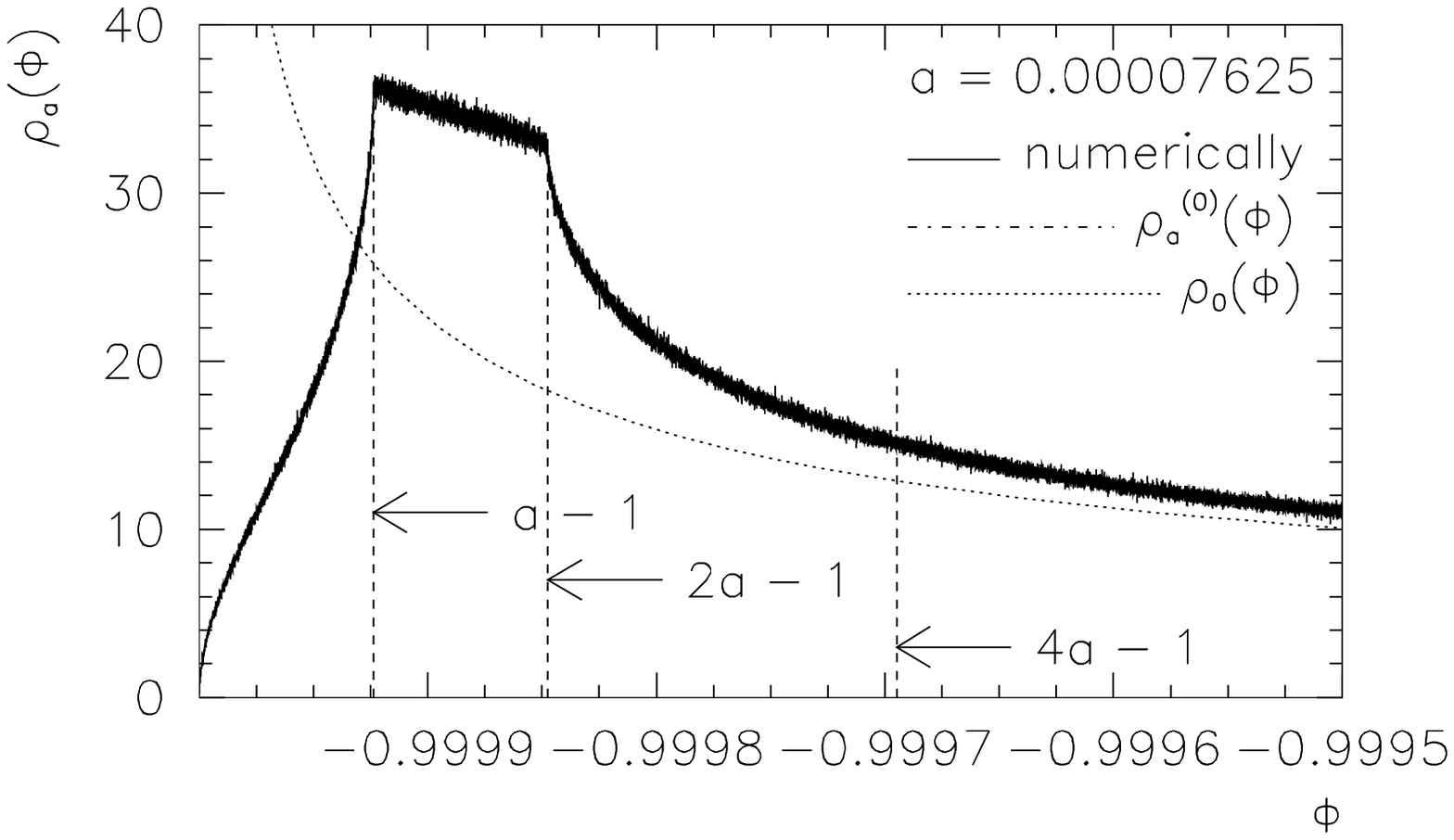, scale=0.8}
\caption{\label{rho23b}Invariant density $\rho_a(\phi)$ close to $\phi=-1$ for
  the 2B-string with $a=0.00007625$, as compared to the invariant
  density $\rho_0(\phi)$ for vanishing coupling (dashed curve)}
\vspace{24pt} \epsfig{figure=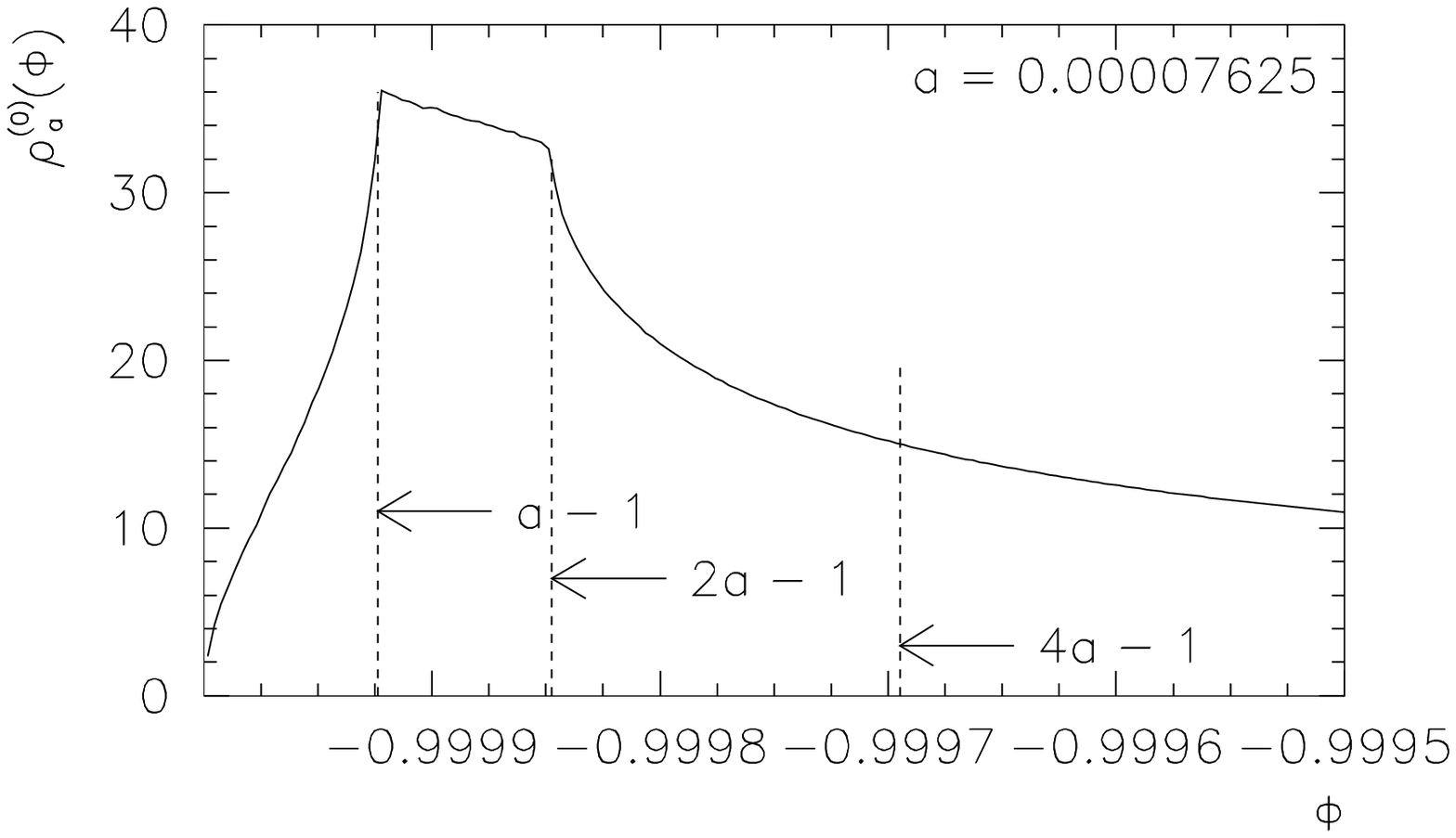, scale=0.8}
\caption{\label{sum23b}Zeroth iterate $\rho_a^{(0)}(\phi)$ close to $\phi=-1$,
  calculated according to Eq.~(\ref{rhoay0sum})}
\end{center}\end{figure}

\subsection{The zeroth iterate of the convolution formula}
Having the density $\rho_{aa}(\tilde\phi)$ at hand, we can now calculate the
probability that the value $\phi$ in an interval of length $d\phi$ and the
value $\tilde\phi$ in an interval of length $d\tilde\phi$ give rise to a value
$\phi'=(1-a)T_2(\phi)+a\tilde\phi$ in an interval of length $d\phi'$,
\begin{equation}\label{rhoaphip}
\rho_a(\phi')=\frac1a\int_{-1}^{+1}\rho_a(\phi)
  \rho_{aa}\left((\phi'-(1-a)T_2(\phi))/a\right)d\phi.
\end{equation}
Note that there is an implicit constraint in this equation that restricts the
integration range of $\phi$, given by the condition $|\tilde{\phi}|\leq 1$. So
far the equation is quite generally valid. Now let us assume that we are close
to the left edge, by writing $\phi'=ay-1$. In this case the integration range
for $\phi$ is restricted by the demand $|\tilde{\phi}|\leq 1$ to a small
symmetric interval around 0. We may thus write $\rho_a(\phi)\approx\rho_a(0)$
to obtain the leading order approximation of the 1-point density
$\rho_a(ay-1)$ close to $\phi=-1$,
\begin{eqnarray}
\rho_a^{(0)}(ay-1)&=&\frac{\rho_a(0)}a\int_{-1}^1
  \rho_{aa}\left((ay-1-(1-a)T_2(\phi))/a\right)d\phi\ =\nonumber\\
  &=&\frac{2\rho_a(0)}a\int_0^1
  \rho_{aa}\left((ay-1-(1-a)T_2(\phi))/a\right)d\phi
\end{eqnarray}
which we call `zeroth iterate' (of our convolution scheme). Finally, we can
change to an integration over $\tilde\phi$,
\begin{eqnarray}
\tilde\phi=\frac1a\left(ay-1-(1-a)(2\phi^2-1)\right)&\Leftrightarrow&
  \phi=\frac1{\sqrt{2(1-a)}}\sqrt{a(y-1-\tilde\phi)}\nonumber\\
  &&d\phi=\frac{-a\,d\tilde\phi}{2\sqrt{2a(1-a)}}\frac1{\sqrt{y-1-\tilde\phi}}
\end{eqnarray}
to obtain
\begin{equation}\label{rhoay0}
\rho_a^{(0)}(ay-1)=\frac{\rho_a(0)}{\sqrt{2a(1-a)}}\int_{-1}^{y-1}
  \frac{\rho_{aa}(\tilde\phi)d\tilde\phi}{\sqrt{y-1-\tilde\phi}}.
\end{equation}
Note that the integration range in this formula is given by the domain of the
square root appearing in the denominator and the demand $|\tilde\phi|\le1$. In
the following, we no longer write down the integration limits explicitly.
Using Eq.~(\ref{rho00}) for the density of $\tilde{\phi}$, we can substitute
$\phi_+=2\tilde\phi-\phi_-$ to end up with the symmetric result
\begin{equation}\label{rhoay0sym}
\rho_a^{(0)}(ay-1)=\frac{\rho_a(0)}{\sqrt{2a(1-a)}}\int
  \frac{\rho_a(\phi_+)d\phi_+\rho_a(\phi_-)
  d\phi_-}{\sqrt{y-1+(\phi_++\phi_-)/2}}.
\end{equation}
The two integration limits in this formula are given by the condition that the
argument of the square root should always be positive and that
$\phi_+,\phi_-\in[-1,1]$. Note that in calculating the leading order
approximation of the density close to $\phi=-1$ there is no need to keep the
indices $a$ on the right hand side of Eq.~(\ref{rhoay0sym}). Therefore, we can
use the explicit representations in Eqs.~(\ref{rho0}) and~(\ref{rho00}) to
calculate the integrals. Another possibility is to calculate the integrals as
time averages. We may just iterate with uncoupled Chebyshev maps $T_2$ to
obtain a  weighting with $\rho_0$. From this we obtain
\begin{equation}\label{rhoay0sum}
\rho_a^{(0)}(ay-1)=\frac1{\pi\sqrt{2a(1-a)}J^2}\sum_{n_+,n_-}
  \frac{\theta(y-1+(\phi_{n_+}+\phi_{n_-})/2)}{\sqrt{y-1
  +(\phi_{n_+}+\phi_{n_-})/2}}
\end{equation}
where $J$ is the total number of iterations. The analytic perturbative result
calculated in this way is shown in Fig.~\ref{sum23b}. It coincides quite
perfectly with the numerically obtained density (obtained by direct iteration
of the CML) as shown in Fig.~\ref{rho23b}.

The main conclusions that we can draw up to now are:
\begin{itemize}
\item The behaviour at the boundary close to $\phi=-1$ is reproduced by the
  zeroth iterate of our convolution scheme with high accuracy. To leading
  order in $a$ we have $\rho_a(ay-1)=\rho_a^{(0)}(ay-1)$.
\item For arbitrarily small $a$ the density $\rho_a(\phi)$ does not approach
  infinity for $\phi\to -1$ (as in the uncoupled case) but falls off to $0$
  after a maximum at $\phi=a-1$.
\item If we parametrize the region close to $\phi=-1$ by $\phi=ay-1$, the only
  dependence on the coupling $a$ that is left for the density is
  $\rho_a(ay-1)\sim a^{-1/2}$.
\end{itemize}
Actually, the last property is already a property of $\rho_0(\phi)$,
\begin{equation}
\rho_0(ay-1)=\frac1{\pi\sqrt{1-(ay-1)^2}}\approx\frac1{\pi\sqrt{2ay}}.
\end{equation}
However, in combination with the appearence of a last maximum at $\phi=a-1$,
we obtain a scaling relation between $a$ and $\phi$, which induces a variety
of interesting results (see section 6).

\subsection{Two-point function and backstep transformation}
\begin{figure}\begin{center}
\epsfig{figure=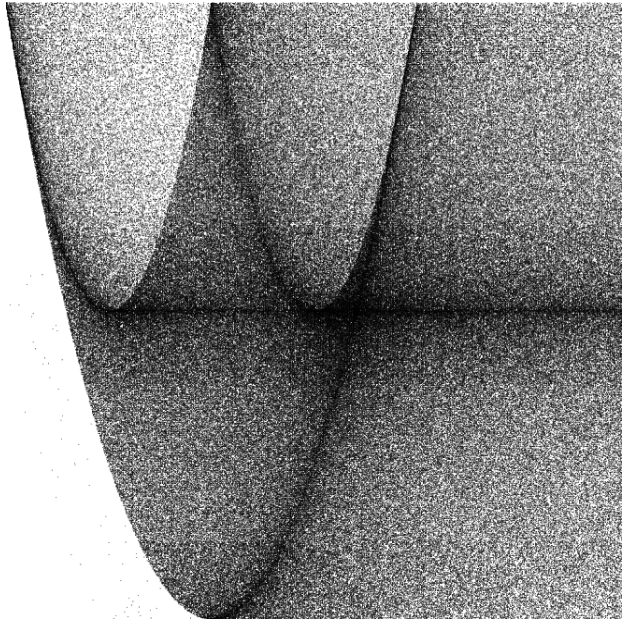, scale=0.5}
\caption{\label{corr23c}Two-point density function
$\rho_{aa}(ay-1,\tilde\phi)$ in the range\hfil\break
$0\le y\le 3$ (horizontally) und $-1\le\tilde\phi\le+1$ (vertically).}
\end{center}\end{figure}

The next iteration step of our scheme leads from the region just considered to
the main scaling region close to $\phi=+1$ which we parametrize by $\phi=1-ax$.
In order to accomplish this step we could think of again calculating the
convolution according to Eq.~(\ref{rhoaphip}), only that in this case we use
$\phi'=1-ax$. However, for values $\phi$ close to the boundaries neighboured
field variables exhibit strong correlations, they cannot be regarded as nearly
independent anymore. The 2-point density function describing the joint
probability of neighboured lattice sites exhibits a rich structure, as shown
in Fig.~\ref{corr23c}.
This structure at the border of the interval represents the main difficulty to
analyse the invariant densities of nonhyperbolic CMLs. Our basic idea in the
following is to proceed by one backward iteration to the region close to
$\phi=0$ where approximate independence holds, and then transform the
densities accordingly.

In order to construct the two-point function analytically, we make use of the
fact that the integral of this function $\rho_{aa}(ay-1,\tilde\phi)$ over
$\tilde\phi$ has to reproduce $\rho_a(ay-1)$ since this is just the marginal
distribution. In searching for a transformation which keeps the integral in
Eq.~(\ref{rhoay0sym}) invariant but changes the integrand we have to think
about an appropriate substitution. The one which fits in for our purposes is
the transformation $\phi_\pm=T_2^{-1}(\phi'_\pm)$ which we may call the
backstep (or pre-image) transformation. The backstep transformation keeps the
unperturbed density invariant,
\begin{equation}
\rho_0(\phi'_\pm)d\phi'_\pm=\rho_0(T_N(\phi_\pm))dT_N(\phi_\pm)
  =N\rho_0(\phi_\pm)d\phi_\pm.
\end{equation}
Under this transformation we obtain quite generally
\begin{equation}
\rho_a^{(0)}(ay-1)=\frac1{\pi\sqrt{2a(1-a)}N^2}\sum^N\sum^N
  \int\frac{\rho_0(\phi'_+)d\phi'_+\rho_0(\phi'_-)d\phi'_-}{\sqrt{y-1
  +(T_N^{-1}(\phi'_+)+T_N^{-1}(\phi'_-))/2}}
\end{equation}
where the two-fold sum runs over all $N$ solutions of the inverse map
$T_N^{-1}(\phi)$. For $N=2$, we sum over the two branches of each of the two
square roots. In substituting finally $\phi'_+=2\tilde\phi-\phi'_-$ we obtain
$\rho_a^{(0)}(ay-1)=\int\rho_{aa}^{(0)}(ay-1,\tilde\phi)d\tilde\phi$ where
\begin{equation}
\rho_{aa}^{(0)}(ay-1,\tilde\phi)=\frac1{\pi\sqrt{2a(1-a)}2^2}\sum_\pm\sum_\pm
  \int\frac{2\rho_0(2\tilde\phi-\phi'_-)\rho_0(\phi'_-)d\phi'_-}
  {\sqrt{y-1+(\pm\sqrt{\tilde\phi+(1-\phi'_-)/2}\pm\sqrt{(1+\phi'_-)/2})/2}}.
\end{equation}
This perturbative result reproduces the numerical 2-point distribution
observed in Fig.~\ref{corr23c}. Finally, as a one-point distribution the
function $\rho_{aa}^{(0)}(ay-1,\tilde\phi)$ has to be normalised. Actually, we
obtain this normalisation by no effort because obviously the ratio
$\hat\rho_{aa}^{(0)}(ay-1,\tilde\phi)=\rho_{aa}^{(0)}(ay-1,\tilde\phi)/
\rho_a^{(0)}(ay-1)$ (describing a conditional probability) is normalised.

\subsection{The first iteration step: from $\phi=-1$ to $\phi=+1$}
The normalisation is skipped again if we use the conditional probability
$\hat\rho_{aa}^{(0)}(ay-1,\tilde\phi)$ close to $\phi=-1$ for the calculation
of the next iterate,
\begin{eqnarray}
\rho_a^{(1)}(\phi')&=&\int\hat\rho_{aa}^{(0)}
  \left(ay-1,(\phi'-(1-a)T_2(ay-1))/a\right)\rho_a^{(0)}(ay-1)dy\ =\nonumber\\
  &=&\int\rho_{aa}^{(0)}\left(ay-1,(\phi'-(1-a)T_2(ay-1))/a\right)dy.
\end{eqnarray}
For $\phi'=1-ax$ the second argument reads $\tilde\phi=1+4y-x+O(a)$. Because
of this, the range of integration for $y$ does not exceed $(1-x)/4$ by more
than one fourth. Instead of integrating over $y$, we can integrate over
$\tilde\phi$ which enables us to reverse the backstep transformation,
\begin{eqnarray}\label{rhoax1}
\lefteqn{\rho_a^{(1)}(1-ax)=\frac14\int\rho_{aa}\left(\frac a4
  (x-1+\tilde\phi)-1,\tilde\phi\right)d\tilde\phi\ =}\nonumber\\
  &=&\frac1{4\pi\sqrt{2a(1-a)}2^2}\sum^2\sum^2
  \int\frac{2\rho_0(2\tilde\phi-\phi'_-)\rho_0(\phi'_-)d\tilde\phi\,
  d\phi'_-}{\sqrt{(x-1+\tilde\phi)/4-1+(T_2^{-1}(2\tilde\phi-\phi'_-)
  +T_2^{-1}(\phi'_-))/2}}\ =\nonumber\\
  &=&\frac1{4\pi\sqrt{2a(1-a)}2^2}\sum^2\sum^2
  \int\frac{\rho_0(\phi'_+)d\phi'_+\rho_0(\phi'_-)d\phi'_-}{\sqrt{(x-1
  +(\phi'_++\phi'_-)/2)/4-1+(T_2^{-1}(\phi'_+)
  +T_2^{-1}(\phi'_-))/2}}\nonumber\\
  &=&\frac1{4\pi\sqrt{2a(1-a)}}
  \int\frac{\rho_0(\phi_+)d\phi_+\rho_0(\phi_-)d\phi_-}{\sqrt{(x-1
  +(T_2(\phi_+)+T_2(\phi_-))/2)/4-1+(\phi_++\phi_-)/2}}\ =\nonumber\\
  &=&\frac1{4\pi\sqrt{2a(1-a)}}
  \int\frac{\rho_0(\phi_+)d\phi_+\rho_0(\phi_-)d\phi_-}{\sqrt{x/4
  +(\phi_++\phi_--2)/2+(T_2(\phi_+)+T_2(\phi_-)-2)/8}}.
\end{eqnarray}
The sums over the branches disappears again. For the last line we have used an
equivalent representation which will later allow us to unify the results.
Again, we can use a time average over iterates of uncoupled Chebyshev maps
to generate the weighting with the density $\rho_0$ in the above integrals.

\begin{figure}\begin{center}
\epsfig{figure=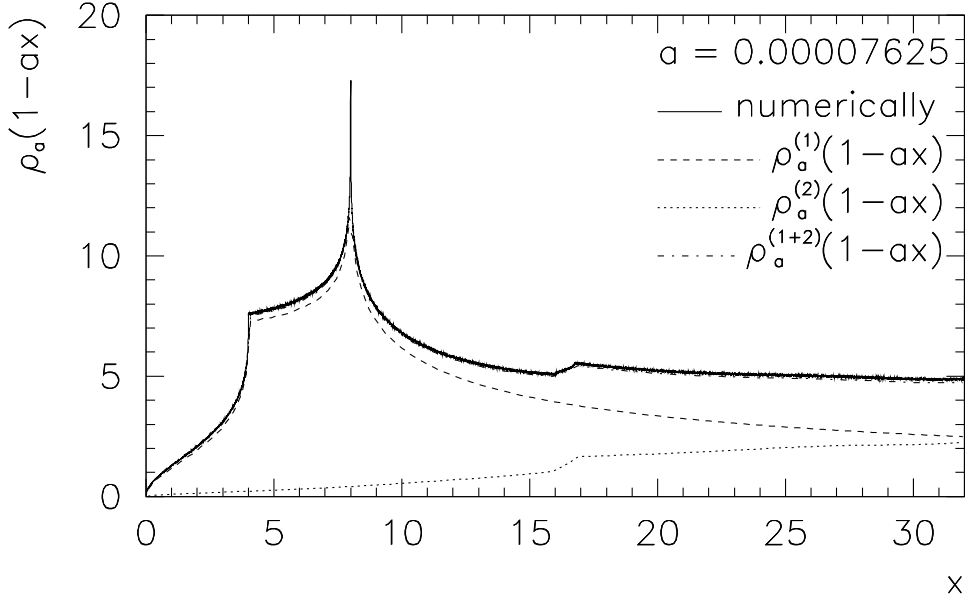, scale=0.8}
\caption{\label{iter2d}1-point density $\rho_a(1-ax)$ at the right edge of the
interval (solid curve) for the 2B-string with $a=0.00006725$ as a function of
$x$, compared to the first iterate $\rho_a^{(1)}(1-ax)$ (dashed curve), the
second iterate $\rho_a^{(2)}(1-ax)$ (dotted curve), and the sum of both
(dashed-dotted curve)}
\end{center}\end{figure}

\subsection{Further iteration steps}
As one can see in Fig.~\ref{iter2d}, the first iterate of our convolution
scheme as given by Eq.~(\ref{rhoax1}) already fits the invariant density
reasonably well. However, it does not reproduce the structure near
$x=16\ldots 17$. Since $T_2$ has an unstable fixed point at $\phi=1$ with
slope +4, this structure is a consequence of local iterates of the CML passing
the previously described correlated border area one step earlier.

In a recursive way, we thus define density contributions (iterates of our
convolution scheme) $\rho_a^{(p)}(1-ax)$ ($p>1$) as follows:
\begin{itemize}
\item perform the backstep transformation
\item calculate the two-point function $\rho_{aa}^{(p-1)}(1-ax',\tilde\phi)$
  by integrating over $\phi'_-$
\item normalise by $\rho_a^{(p-1)}(1-ax')$
\item integrate over $x'$ (the normalisation cancels)
\item retract the backstep transformation
\end{itemize}
Following this recipe, we obtain
\begin{equation}\label{rhoaxp}
\rho_a^{(p)}(1-ax)=\frac1{4^p\pi\sqrt{2a(1-a)}}\int\frac{\rho_0(\phi_+)d\phi_+
  \rho_0(\phi_-)d\phi_-}{\sqrt{x/4^p+r_2^p(\phi_+)+r_2^p(\phi_-)}},\qquad
r_2^p(\phi):=\frac12\sum_{q=0}^p\frac{T_{2^q}(\phi)-1}{4^q}.
\end{equation}
The entire invariant density is given by the sum over all these contributions,
\begin{equation}
\rho_a(1-ax)=\sum_{p=1}^\infty\rho_a^{(p)}(1-ax)
  =\sum_{p=1}^\infty\frac1{4^p\pi\sqrt{2a(1-a)}}\int
  \frac{\rho_0(\phi_+)d\phi_+\rho_0(\phi_-)d\phi_-}{\sqrt{x/4^p
  +r_2^p(\phi_+)+r_2^p(\phi_-)}}.   \label{21}
\end{equation}
The series converges rapidly, so usually it is sufficient to take into account
the first few functions $\rho_a^{(p)}$ only. Roughly speaking, the
contribution $\rho_a^{(p)}$ takes care of those points that passed the
nonhyperbolic region $p$ steps ago.

Apparently Eq.~(\ref{rhoaxp}) does even hold for the case $p=0$, taking into
account that the integral is invariant under the replacements
$\phi_\pm\to-\phi_\pm$. However, in this case the left hand side has to be
$\rho_a^{(0)}(ax-1)$, i.e.\ the zeroth iterative of our scheme located close
to $\phi=-1$.

Fig.~\ref{iter2d} shows our analytic results $\rho_a^{(1)}(1-ax)$,
$\rho_a^{(2)}(1-ax)$ and the sum of these two contributions. There is
excellent agreement with the numerical histogram. The agreement is so good
that the sum of the two terms is not visible behind the numerical data points.

\section{Explaining the selfsimilar shape\\ of the density patterns}
\begin{figure}\begin{center}
\epsfig{figure=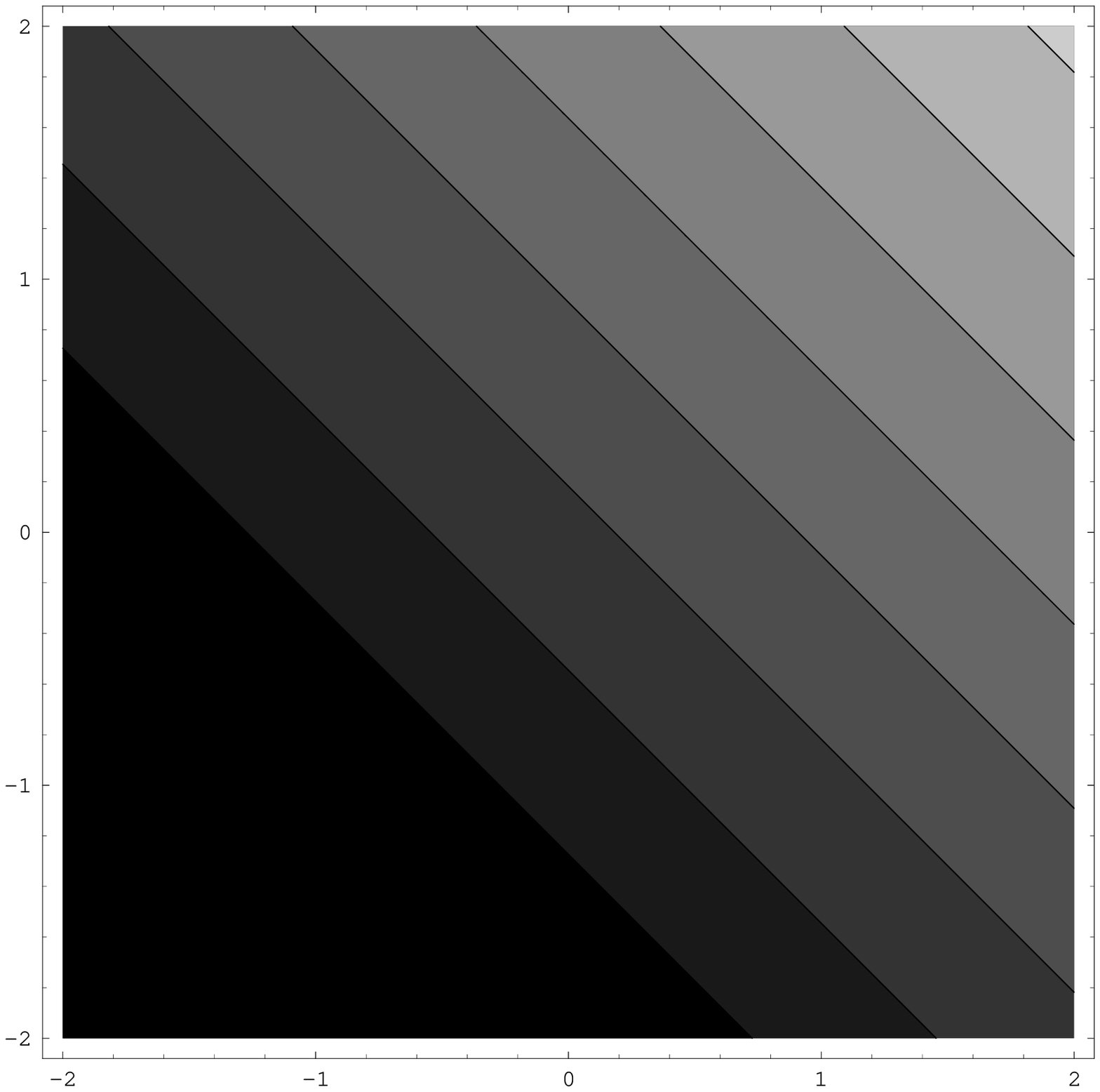, scale=0.4}\qquad
\epsfig{figure=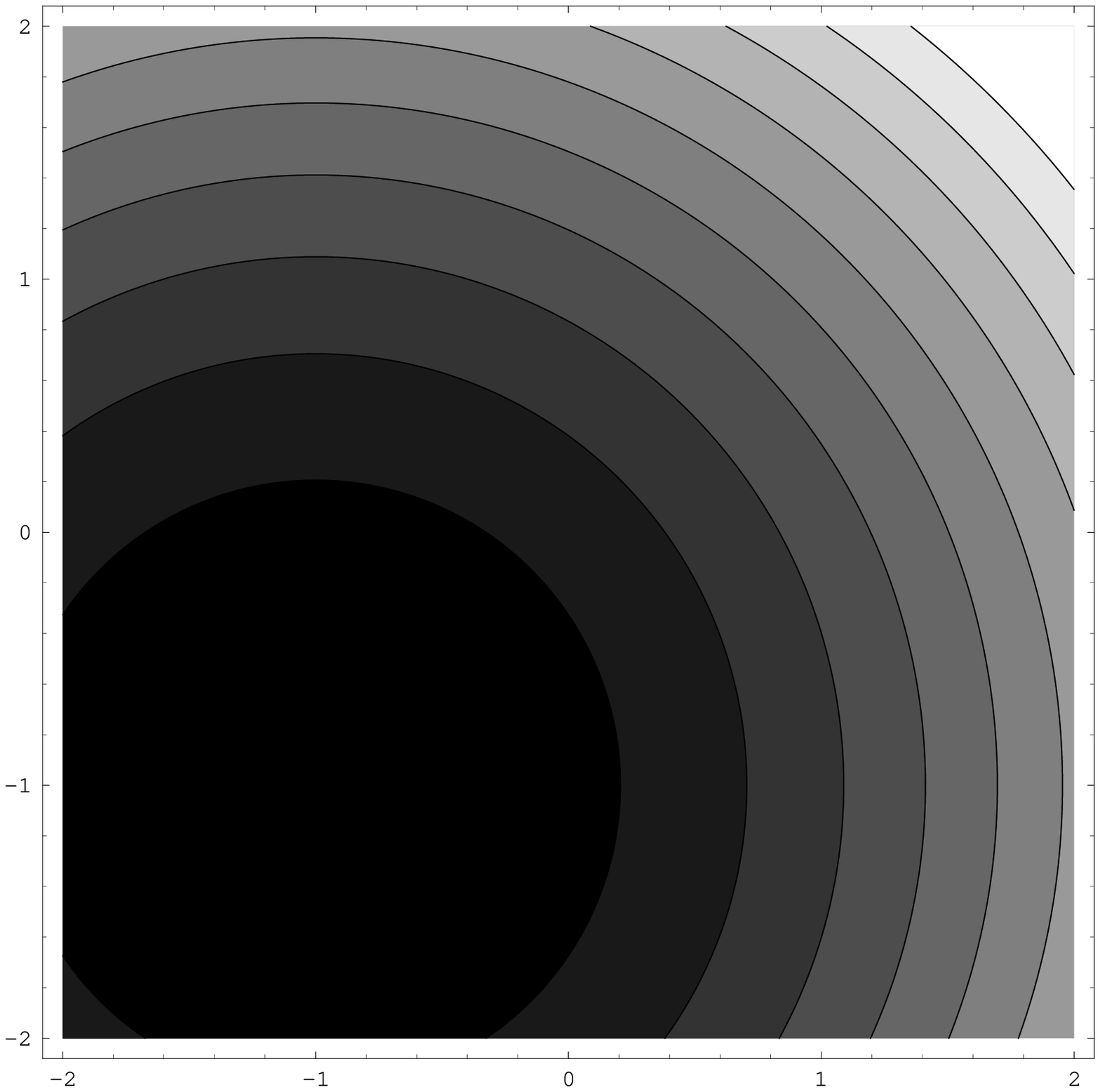, scale=0.4}\\
\centerline{(a)\kern192pt(b)}\vspace{12pt}
\epsfig{figure=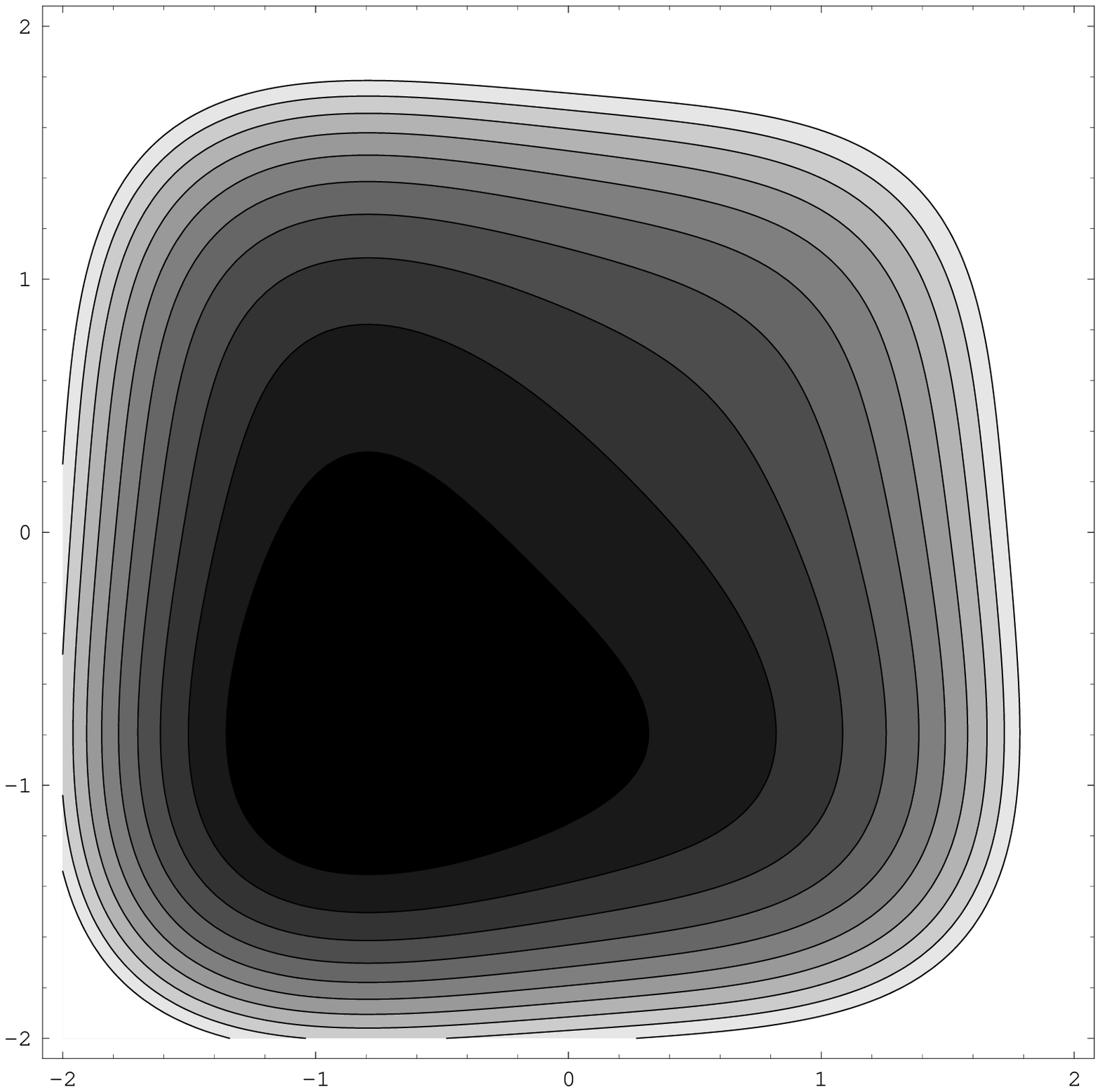, scale=0.4}\qquad
\epsfig{figure=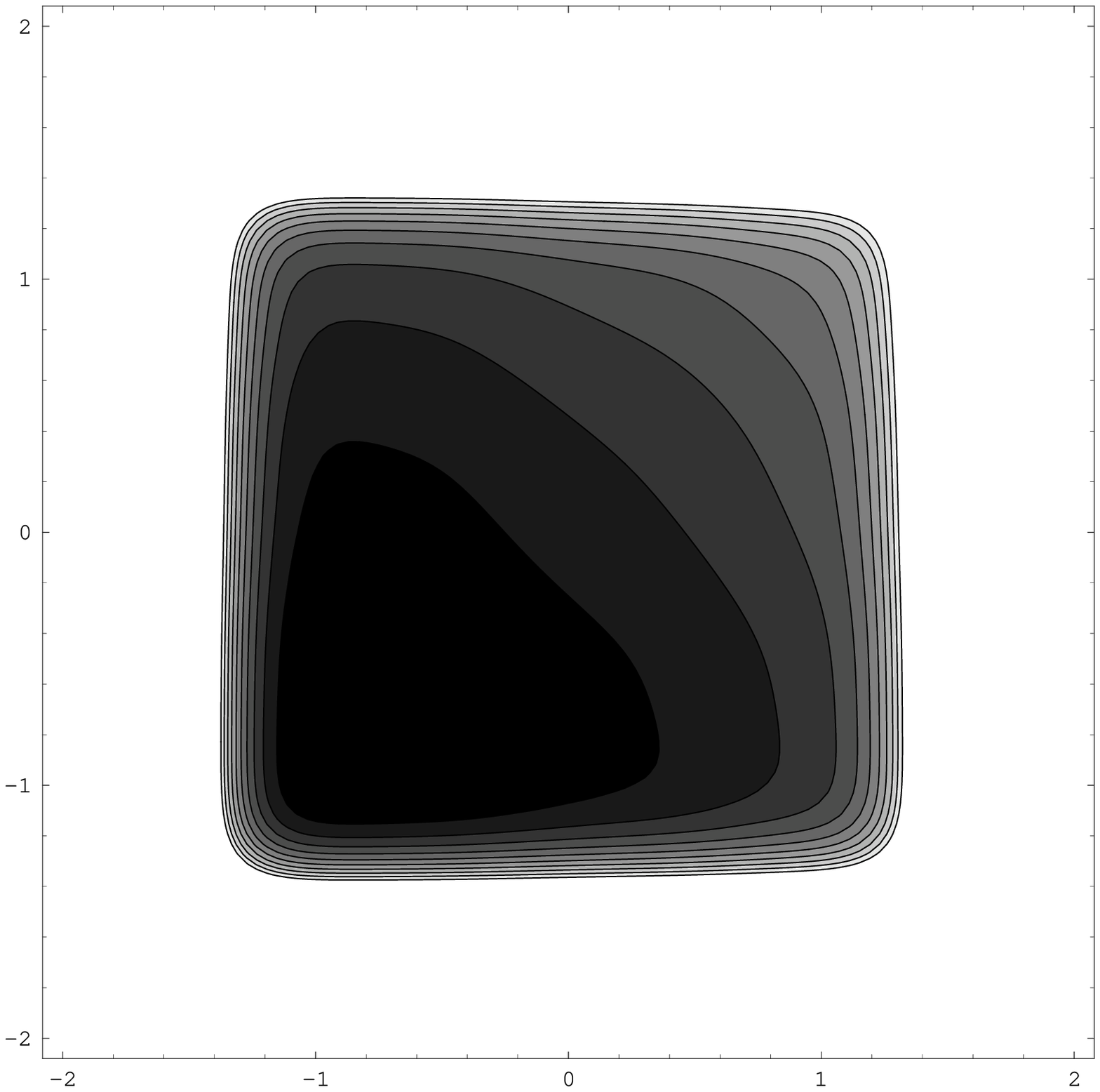, scale=0.4}
\centerline{(c)\kern192pt(d)}
\caption{\label{rad2m}Contour plots of the function
  $r_2^p(\phi_+)+r_2^p(\phi_-)$ for the values $p=0$ (a), $p=1$ (b),
  $p=2$ (c), and $p=3$ (d) in the range given by $\phi_\pm\in[-2,2]$. Dark
  shadings indicate low values, bright shadings high values of the function in
  $[-2,2]$.}
\end{center}\end{figure}

It is obvious that the square root in the denominator of Eq.~(\ref{21}) has
essential influence on the shape of the invariant 1-point density of the CML,
since the argument of the square root needs to be positive which determines
the integration range in our formulas. Abrupt changes of the density pattern
are connected to zero-border crossings of the arguments of the square roots.
Looking more closely at the function
\begin{equation}
r_2^p(\phi_+,\phi_-,x)=\frac x{4^p}+r_2^p(\phi_+)+r_2^p(\phi_-),
\end{equation}
we can find reasons for the patterns. In order to do this analysis, we
generated the contour plots of $r_2^p(\phi_+)+r_2^p(\phi_-)$ for the values
$p=0,1,2,3$ by using MATHEMATICA. These plots are shown in Fig.~\ref{rad2m}.
It is obvious that the demand $r_2^p(\phi_+,\phi_-,x)\ge 0$ describes a
nontrivial limitation of the basic integration range $\phi_\pm\in[-1,1]$. The
equation $r_2^p(\phi_+,\phi_-,x)=0$ for different values of $x$ describes
the analytic curves which appear as contour curves in Fig.~\ref{rad2m}.

\subsection{Patterns generated by $p=0$}
The analytic curve given by the vanishing of the radical
\begin{equation}
r_2^0(\phi_+,\phi_-,y)=y+r_2^0(\phi_+)+r_2^0(\phi_-)=y+(\phi_++\phi_-)/2-1
\end{equation}
describes an off-diagonal straight line with intercept $2(y-1)$ that limits
the integration range to lower values of $\phi_+$ and $\phi_-$ (cf.\
Fig.~\ref{rad2m}(a)). We observe that for $y=0$ the basic integration range is
totally suppressed, the integration area then increases like $(2y)^2/2$ up to
$y=1$ and like $(4+(2-y)^2)/2$ up to $y=2$. Starting from $y=2$, the basic
integration range is no longer limited by the contraint
$r_2^0(\phi_+,\phi_-,y)\ge 0$. In accordance to this, the zeroth-order density
contribution starts with the value zero at $y=0$, shows non-differentiable
cusps at $y=1$ and $y=2$ and monotonic behaviour elsewhere (compare
Fig.~\ref{rho23b}).

\subsection{Patterns generated by $p=1$}
Here we need to analyse the radical
\begin{equation}
r_2^1(\phi_+,\phi_-,x)=\frac x4+r_2^1(\phi_+)+r_2^1(\phi_-)
  =\frac14\left(x+(\phi_++1)^2+(\phi_-+1)^2-8\right),
\end{equation}
For $x<8$ the analytic curve parametrized by $r_2^1(\phi_+,\phi_-,x)=0$ is a
circle in the $(\phi_+,\phi_-)$-plane with radius $\sqrt{8-x}$ and center
located at $(-1,-1)$ (cf.\ Fig.~\ref{rad2m}(b)). The intersection of the
outside of this circle and the inside of the rectangle $-1\le\phi_\pm\le 1$
determines the integration range.
\begin{itemize}
\item For $x=0$ the rectangle is totally covered by the circle. In this case,
  the integration range vanishes. We see that the $p=1$ density contribution
  vanishes for $x=0$ as well.
\item Between $x=0$ and $x=4$ the circle uncovers the two positive edges of
  the rectangle, given by the lines $\phi_\pm=+1$. We see that the $p=1$
  density contribution grows monotonically in this interval.
\item Between $x=4$ and $x=8$ the circle uncovers also the two negative edges
  of the rectangle, given by the lines $\phi_\pm=-1$. In this interval the
  $p=1$ density contribution declines.
\item Above $x=8$ the whole rectangle is uncovered, so there is no restriction
  of the integration range anymore. Starting from $x=8$ there are no further
  cusps. Caused by the increasing denominator, the $p=1$ density contribution
  declines.
\end{itemize}

\subsection{Patterns generated by $p=2$}
The second iterate $\rho_a^{(2)}(1-ax)$ shown in Fig.~\ref{iter2d} has a cusp
at $x=17$ which is responsible for the shift of the cusp of the invariant
density by an amount of $-\log_4(17/16)$ in Fig.~\ref{ener2n}. In a similar
way as for $p=1$, we can proceed for $p=2$. The only point is that we do no
longer have a name for the algebraic curve described by
$r^2_2(\phi_+,\phi_-,x)=0$. However, the behaviour is similar as the shape of
the curve is a kind of mixture between a circle, a rectangle and a triangle
(cf.\ Fig.~\ref{rad2m}(c)). If we analyse the analytic curve for $p=2$ more
closely, we find that the positive edges of the rectangle constituting the
basic integration range are fully uncovered not at $x=16$ as could be assumed
for the circle but at $x=17$. However, starting from $x=16$ the positive edges
start to be uncovered from the negative side. Therefore, we obtain the two
cusps at $x=16$ and $x=17$ as observed. Finally, the basic integration range
becomes totally uncovered for a value $x\approx 33.524$, giving rise to a last
cusp of this pattern at distance $-\log_4(x/16)\approx-0.534$ from the first
cusp (corresponding to $x=16$).

\subsection{Patterns generated by $p\geq 3$}
\begin{figure}\begin{center}
\epsfig{figure=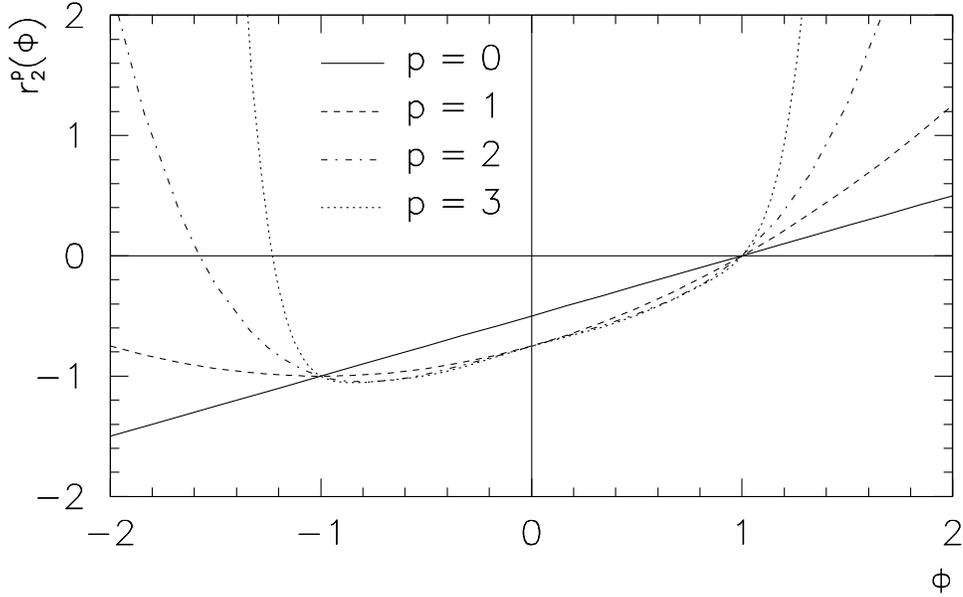, scale=0.8}
\caption{\label{rad2part}the function $r_2^p(\phi)$ for the values
  $p=0,1,2,3$}
\end{center}\end{figure}

If we look at the function $r_2^p(\phi)$ for even higher values of $p$ we make
an iteresting observation. While for the outside region $|\phi|>1$ the
function $r_2^p(\phi)$ starts to diverge, it rapidly converges to a limiting
function inside the region $|\phi|\le 1$. This behaviour is shown in
Fig.~\ref{rad2part}.
Only the inside region is essential for the integration. Therefore, we can
conclude that $r_2^p(\phi)\approx r_2^\infty(\phi)$ for relatively small $p$
and $|\phi|\le 1$. As a consequence of this, there is invariance of the
patterns generated by the higher iterates of our scheme under the replacement
$x\rightarrow x/4$. As a general observation for the radical function
$r_2^p(\phi)$ we add that $r_2^p(\phi=+1)=0$, $r_2^p(\phi=0)=-3/4$ (for $p>0$),
and $r_2^p(-\phi)=r_2^p(\phi)-\phi$ which is caused by the symmetry property
of the Chebyshev polynomial $T_2(\phi)$ and its iterates
$T_{2^q}(\phi)=T_2(T_{2^{q-1}}(\phi))$.

\section{The scaling behaviour for coupled\\ 3rd order Chebyshev maps}
All the steps we have performed in order to explain the scaling behaviour of
the 2B-string can be performed for the 3B-string (coupled 3rd-order
Chebyshev maps with $s=1,b=0$) as well. Scaling regions of the density are
transferred by convolution,
\begin{equation}\label{convol}
\rho_a(\phi')
  =\frac1a\int\rho_a(\phi)\rho_{aa}\left((\phi'-(1-a)T_3(\phi))/a\right)d\phi.
\end{equation}
However, in order to integrate over the argument $\tilde\phi$ of the density
$\rho_{aa}(\tilde\phi)$ we have to solve the cubic equation
\begin{equation}
T_3(\phi)=4\phi^3-3\phi=\frac{\phi'-a\tilde\phi}{1-a}=:\hat\phi.
\end{equation}
The three real solutions are
\begin{eqnarray}
\phi_+&=&\cos\left(\frac13\arccos\hat\phi\right),\qquad
  +\frac12\le\phi_0\le+1\nonumber\\
\phi_0&=&\cos\left(\frac13\arccos\hat\phi+\frac{2\pi}3\right),\qquad
  -\frac12\le\phi\le+\frac12\nonumber\\
\phi_-&=&\cos\left(\frac13\arccos\hat\phi+\frac{4\pi}3\right),\qquad
  -1\le\phi\le-\frac12
\end{eqnarray}
describing the three branches of the inverse map $T_3^{-1}(\hat\phi)$ (similar
formulas apply, of course, to the $N$-th order Chebyshev map, where there
are $N$ real solutions). In order to replace the differential $d\phi$ in
Eq.~(\ref{convol}) by $d\tilde\phi$, we have to specify which of the branches
are involved. The demand $|\tilde\phi|\le 1$ leads to
\begin{equation}
\frac{\phi'-a}{1-a}\le\hat\phi\le\frac{\phi'+a}{1-a}.
\end{equation}

\subsection{Iterates close to $\phi=+1$}
If we use $\phi'=1-ax$, $\hat\phi$ is restricted by
\begin{equation}
1-ax\approx\frac{1-a(1+x)}{1-a}\le\hat\phi\le\frac{1+a(1-x)}{1-a}
  \approx 1+a(2-x)
\end{equation}
and by $-1\le\hat\phi\le+1$. $\hat\phi$ turns out to be almost $+1$, its
pre-images under the Chebyshev map $\hat\phi=T_3(\phi)$ are found at
$\phi=+1$ and $\phi=-1/2$. The unperturbed starting contribution of our
convolution scheme, therefore, comes from $\phi=-1/2$, involving the branches
$\phi_0$ and $\phi_-$. We obtain (for $\hat\phi\approx 1$)
\begin{eqnarray}
d\phi_0&=&-\sin\left(\frac13\arccos\hat\phi+\frac{2\pi}3\right)
  \frac{-d\hat\phi}{3\sqrt{1-\hat\phi^2}}
  \ \approx\ \frac{d\hat\phi}{2\sqrt{3(1-\hat\phi^2)}},\\
d\phi_-&=&-\sin\left(\frac13\arccos\hat\phi+\frac{4\pi}3\right)
  \frac{-d\hat\phi}{3\sqrt{1-\hat\phi^2}}
  \ \approx\ \frac{-d\hat\phi}{2\sqrt{3(1-\hat\phi^2)}}.
\end{eqnarray}
Starting with the integration range for $\hat\phi$ as indicated before, both
integrations over $\phi_0$ and $\phi_-$ are mapped onto this integration range.
If we take into account that the integration range for $\phi_0$ declines from
$\phi=-1/2$ to lower values, we can combine both $d\phi_0$ and $d\phi_-$ to
obtain
\begin{equation}
d\phi=\frac{-d\hat\phi}{\sqrt{3(1-\hat\phi^2)}}
  =\frac{a\,d\tilde\phi}{(1-a)\sqrt{3(1-(1-ax-a\tilde\phi)^2/(1-a)^2)}}
  \approx\frac{a\,d\tilde\phi}{\sqrt{6a(x-1+\tilde\phi)}}
\end{equation}
where we inserted $\hat\phi$ for $\phi'=1-ax$ and calculated the leading order
approximation in $a$. Again, the right hand side of Eq.~(\ref{convol}) needs
to be calculated only to leading order in $a$. Therefore, we replace all
density indices on this side by $0$. Finally, as $\phi$ is close to
$\phi=-1/2$, we can replace $\rho_0(\phi)$ by the constant value
$\rho_0(-1/2)=2/\pi\sqrt3$ to obtain the zeroth iterate
\begin{eqnarray}
\rho_a^{(0)}(1-ax)&=&\frac2{3\pi\sqrt{2a}}\int\frac{\rho_{00}(\tilde\phi)
  d\tilde\phi}{\sqrt{x-1+\tilde\phi}}
  \ =\ \frac2{3\pi\sqrt{2a}}\int\frac{2\rho_0(2\tilde\phi-\phi_-)\rho_0(\phi_-)
  d\tilde\phi d\phi_-}{\sqrt{x-1+\tilde\phi}}\ =\nonumber\\
  &=&\frac2{3\pi\sqrt{2a}}\int\frac{\rho_0(\phi_+)d\phi_+\rho_0(\phi_-)
  d\phi_-}{\sqrt{x-1+(\phi_++\phi_-)/2}}.\qquad(\phi_+=2\tilde\phi-\phi_-)
\end{eqnarray}
The zeroth-order two-point density function $\rho_{aa}^{(0)}(\phi,\tilde\phi)$
is obtained by the backstep transformation
\begin{equation}
\rho_a^{(0)}(1-ax)=\frac2{3\pi\sqrt{2a}}\ \frac1{3^2}\sum^3\sum^3
  \int\frac{\rho_0(\phi'_+)d\phi'_+\rho_0(\phi'_-)d\phi'_-}{\sqrt{x-1
  +(T_3^{-1}(\phi'_+)+T_3^{-1}(\phi'_-))/2}}.
\end{equation}
The first iterate of our convolution scheme is again the integral of the
two-point function,
\begin{equation}
\rho_a^{(1)}(1-ax)=\int\rho_{aa}^{(0)}(1-ax',\tilde\phi')dx'
\end{equation}
where
\begin{equation}
\tilde\phi'=\frac1a\left((1-ax)-(1-a)T_3(1-ax')\right)\approx 9x'+1-x.
\end{equation}
\begin{figure}\begin{center}
\epsfig{figure=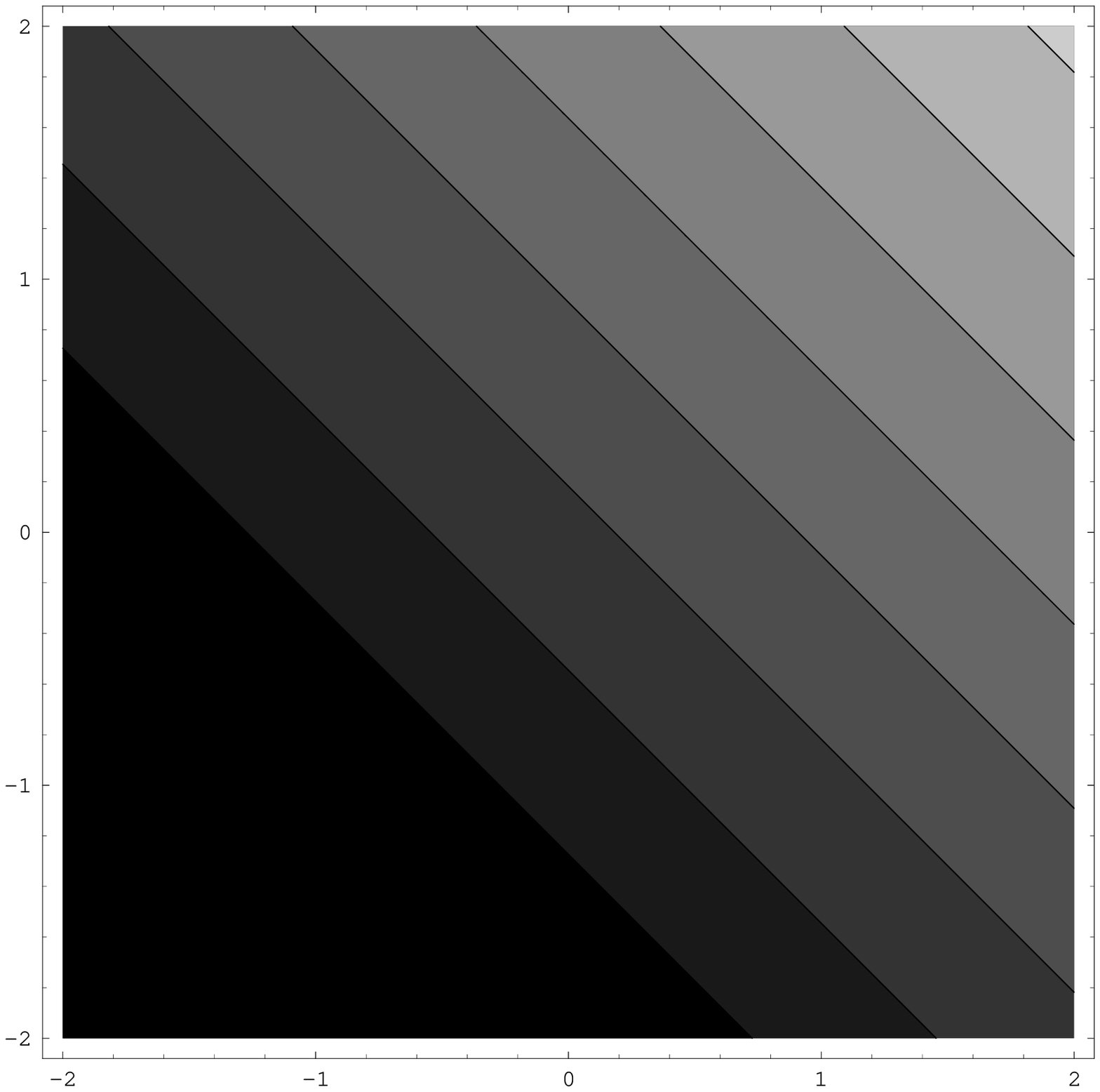, scale=0.4}\qquad
\epsfig{figure=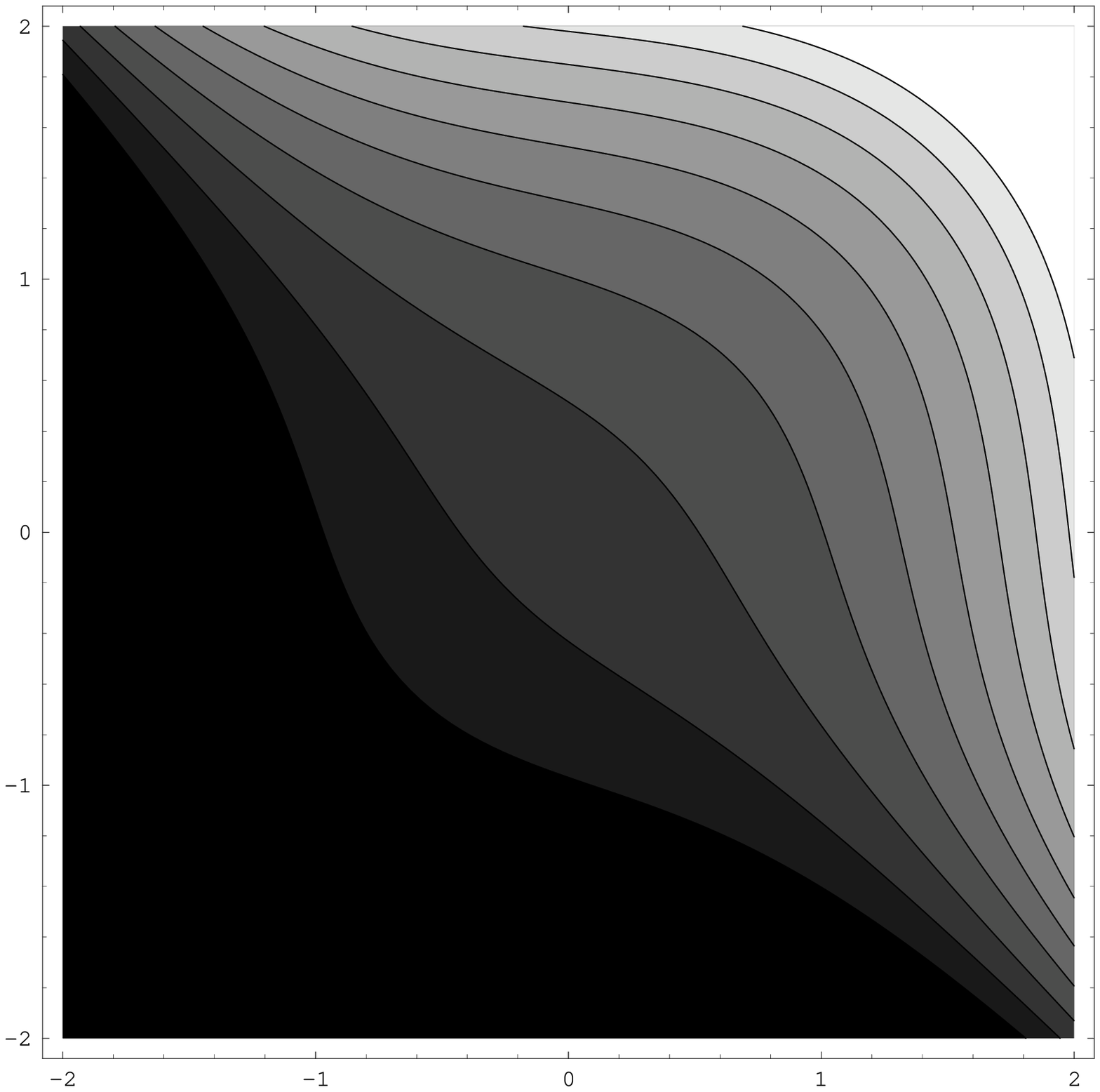, scale=0.4}\\
\centerline{(a)\kern192pt(b)}\vspace{12pt}
\epsfig{figure=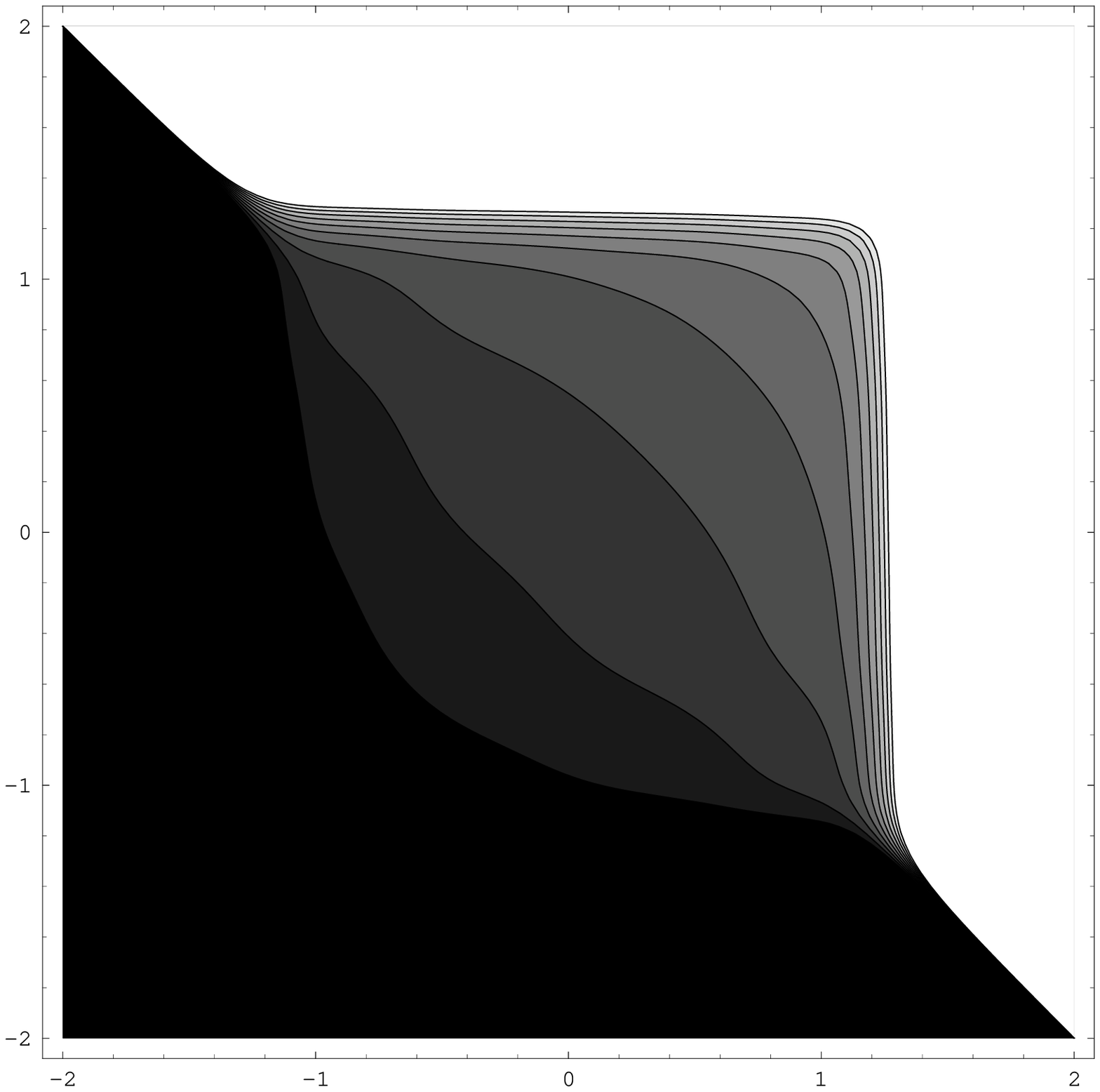, scale=0.4}\qquad
\epsfig{figure=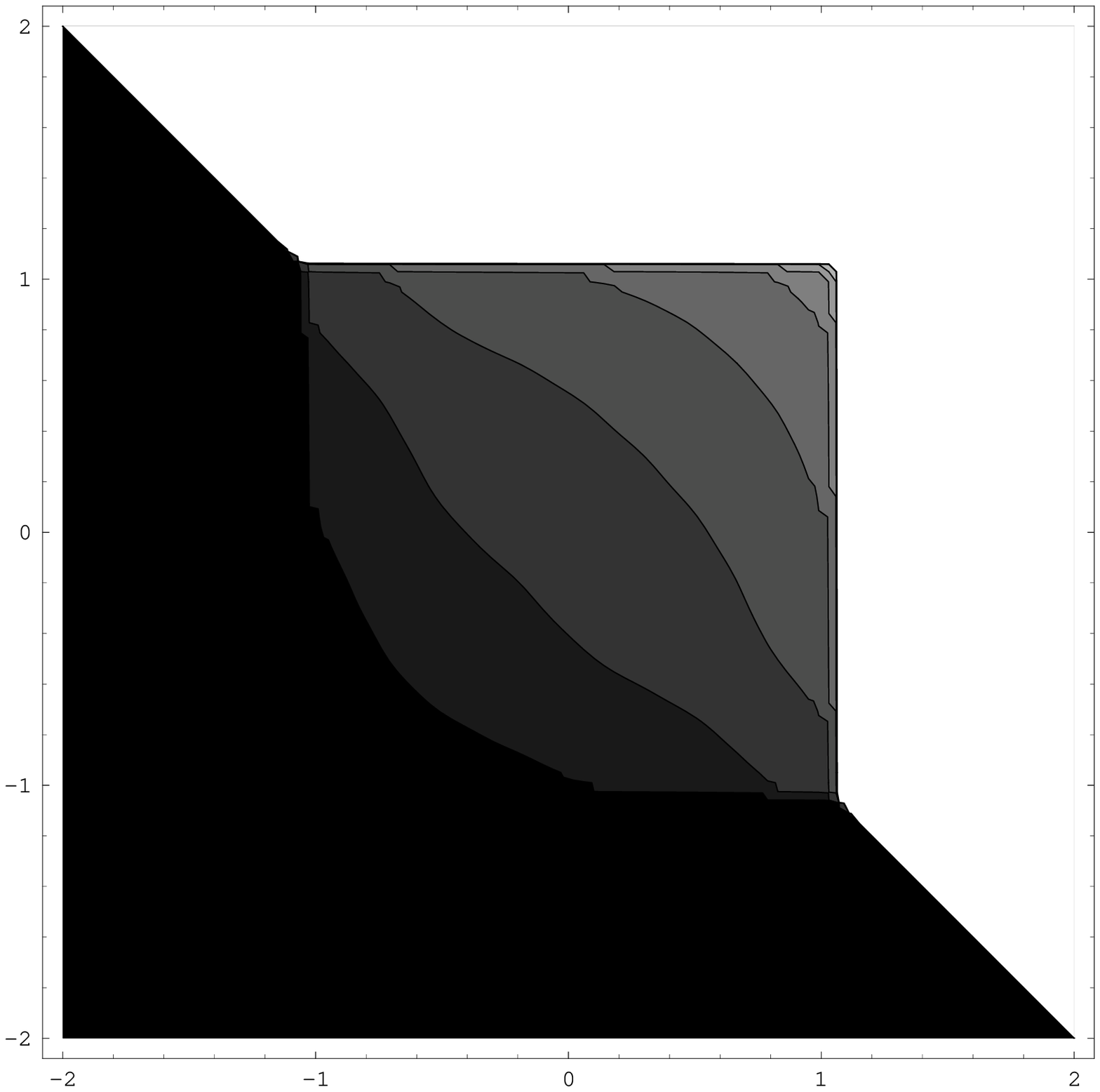, scale=0.4}
\centerline{(c)\kern192pt(d)}
\caption{\label{rad3}Contour plots of the function
  $r_3^p(\phi_+)+r_3^p(\phi_-)$ for the values $p=0$ (a), $p=1$ (b), $p=2$ (c),
  and $p=3$ (d) in the range given by $\phi_\pm\in[-2,2]$. Dark shadings
  indicate low values, bright shadings high values of the function in $[-2,2]$}
\end{center}\end{figure}
Integrating over $\tilde\phi'$ instead of $x'$ and substituting
$\tilde\phi'=(\phi'_++\phi'_-)/2$, we can again revert the backstep
transformation to obtain the first iterate
\begin{equation}
\rho_a^{(1)}(1-ax)=\frac2{27\pi\sqrt{2a}}\int\frac{\rho_0(\phi_+)d\phi_+
  \rho_0(\phi_-)d\phi_-}{\sqrt{x/9+(\phi_++\phi_--2)/2
  +(T_3(\phi_+)+T_3(\phi_-)-2)/18}}.
\end{equation}
Continuing in the same manner, we obtain
\begin{equation}
\rho_a(1-ax)=\sum_{p=0}^\infty\rho_a^{(p)}(1-ax)
  =\sum_{p=0}^\infty\frac2{9^p3\pi\sqrt{2a}}\int\frac{\rho_0(\phi_+)d\phi_+
  \rho_0(\phi_-)d\phi_-}{\sqrt{x/9^p+r_3^p(\phi_+)+r_3^p(\phi_-)}}
\end{equation}
where
\begin{equation}
r_3^p(\phi)=\frac12\sum_{q=0}^p\frac{T_{3^q}(\phi)-1}{9^q}.
\end{equation}

\begin{figure}\begin{center}
\epsfig{figure=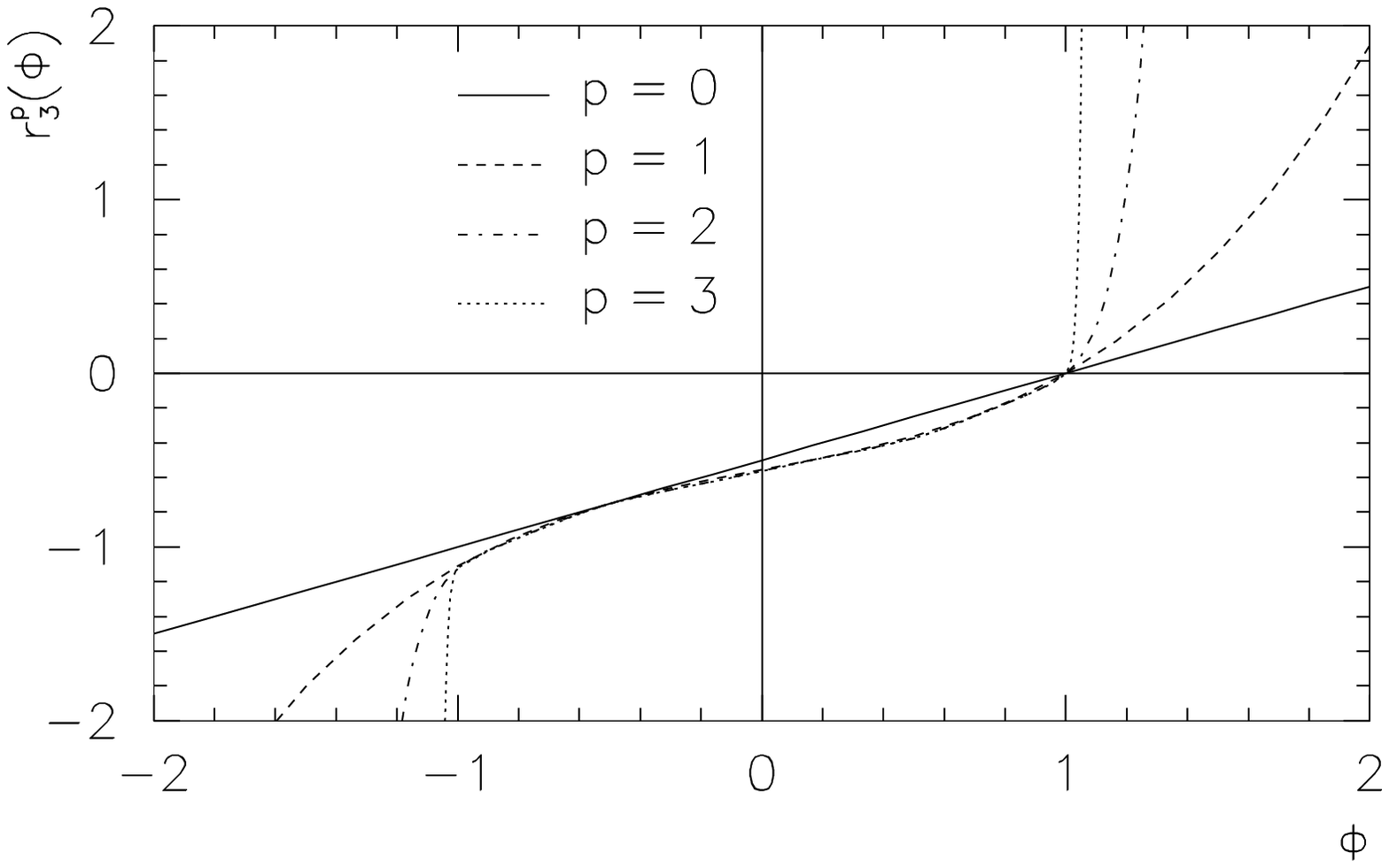, scale=0.8}
\caption{\label{rad3part}the function $r_3^p(\phi)$ for the values
  $p=0,1,2,3$}
\end{center}\end{figure}

\subsection{Iterates close to $\phi=-1$}
We no longer have to go into details for the calculation close to the lower
boundary $\phi=-1$. The unperturbed domain for the zeroth iterate is given by
a region close to $\phi=+1/2$, the higher iterates remain close to $\phi=-1$.
The final result reads
\begin{equation}
\rho_a(ay-1)=\sum_{p=0}^\infty\rho_a^{(p)}(ay-1)
  =\sum_{p=0}^\infty\frac2{9^p3\pi\sqrt{2a}}\int\frac{\rho_0(\phi_+)d\phi_+
  \rho_0(\phi_-)d\phi_-}{\sqrt{y/9^p+r_3^p(\phi_+)+r_3^p(\phi_-)}}.
\end{equation}
We obtain $\rho_a(ay-1)=\rho_a(1-ay)$, i.e.\ there is symmetry of the 1-point
density as expected from general principles~\cite{Beck:2002}. Finally, we add
the remark that the zeroth iterate for the 3B-string close to the lower
boundary $\phi=-1$ should coincide (up to a general factor) with the zeroth
iterate (and, therefore, with the density) of the 2B-string close to
$\phi=-1$, see section 5.4 for a numerical check of this fact.

\subsection{The 3B-pattern}
What to predict about the 3B-pattern? In Fig.~\ref{rad3} we display the
contour plots of the function $r_3^p(\phi_+)+r_3^p(\phi_-)$ for the values
$p=0,1,2,3$. There is a step in passing the off-diagonal for $|\phi_\pm|>1$,
whose size increases with $p$. However, in the region relevant for the
integration ($-1\le\phi_\pm\le+1$) the function converges. In
Fig.~\ref{rad3part} we show $r_3^p(\phi)$ for the same values $p=0,1,2,3$. Up
to a shift which is given by
\begin{equation}\label{r3p0}
r_3^p(0)=-\frac12\sum_{q=0}^p\frac1{9^q}=-\frac{1-(1/9)^{p+1}}{2(1-1/9)}
\buildrel p\to\infty\over\longrightarrow-\frac9{16},
\end{equation}
the odd symmetry of the Chebyshev polynomial $T_3(\phi)$ and its iterates
$T_{3^q}(\phi)=T_3(T_{3^{q-1}}(\phi))$ cause $r_3^p(\phi)-r_3^p(0)$ to have
odd symmetry. On the other hand, as in case of the 2B-string we have
$r_3^p(\phi=+1)=0$ because of $T_{3^q}(\phi=+1)=+1$. Taking these two facts
into account, we can determine the values of $r_3^p(\phi)$ at $\phi=-1$ to be
two times the value at $\phi=0$. Analysing finally the inequality
$x/9^p+r_3^p(\phi_+)+r_3^p(\phi_-)>0$ that determines the restriction of the
basic integration range $-1\le\phi_\pm\le+1$, we can understand the position
of the cusps of the density curve.
\begin{itemize}
\item For $p=0$ the restriction reads $x-1+(\phi_++\phi_-)/2>0$. The pattern
  turns out to be the same as for the 2B-string. We obtain full cover for
  $x=0$ (i.e. full suppression of the integration range), half cover for
  $x=1$, and the disappearence of the covering triangle from the basic
  (rectangular) integration range for $x=2$. This reproduces the three
  observed segments of the density curve.
\item For $p=1$ the restriction reads
  $x-10+2\phi_+^3+3\phi_++2\phi_-^3+3\phi_->0$. Therefore, we should obtain
  half cover at $x=10$ and disappearence of the cover for $x=20$.
\item For high ($p$-th) iterates, we can apply the limit in Eq.~(\ref{r3p0}).
  We observe that the half cover should be reached for $x=(9/8)9^p$, the
  disappearence of the cover for $x=(9/4)9^p$. All these numbers determine
  non-differentiable points of the density.
\end{itemize}

\begin{figure}\begin{center}
\epsfig{figure=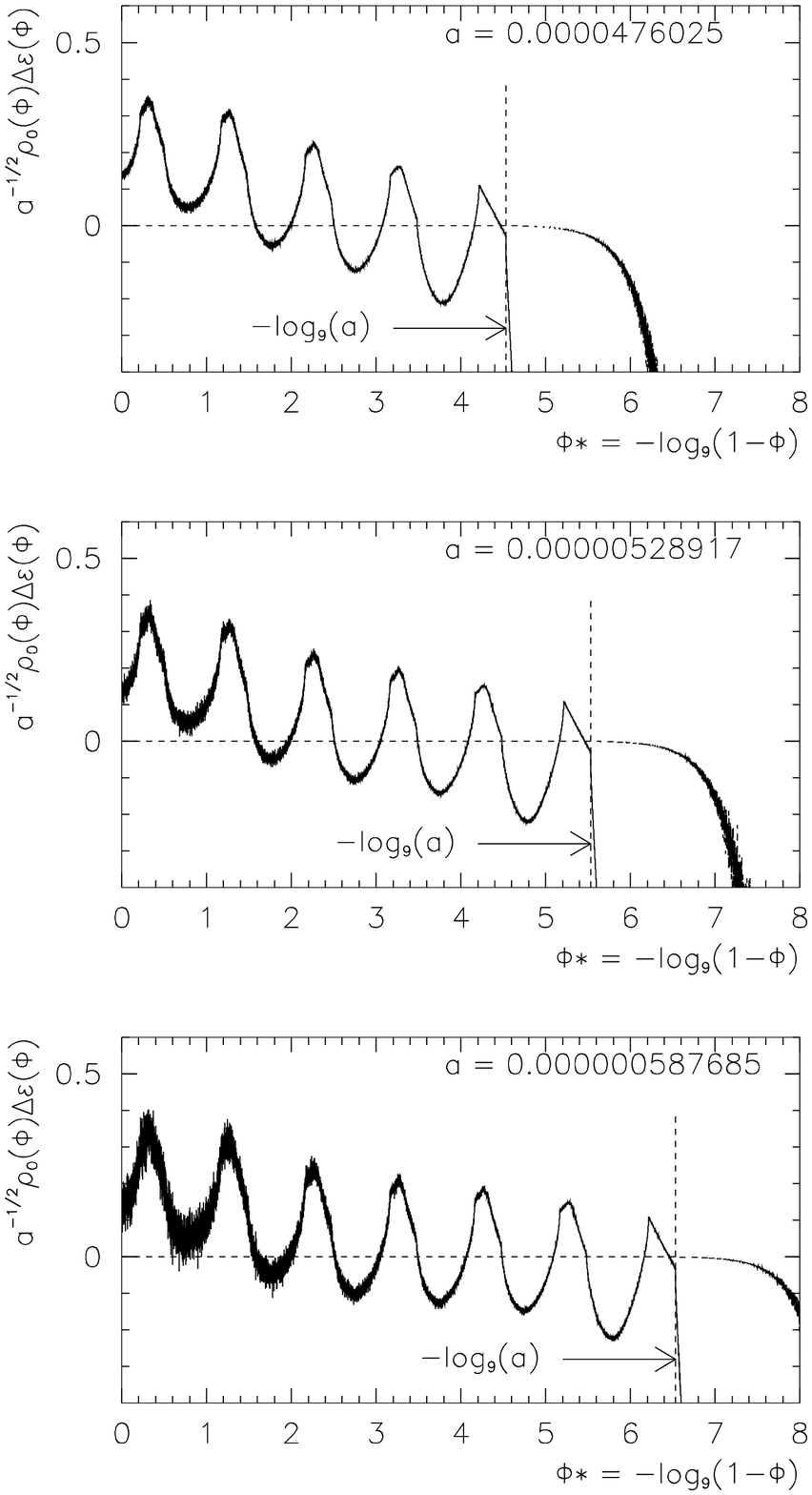, scale=0.75}
\caption{\label{ener3345}Adjusted energy difference
  $a^{-1/2}\rho_0(\phi)\Delta\epsilon(\phi)$ of the 3B-string for
  $a=0.0000476025$, $a=0.00000528917$, and $a=0.000000587685$ (from top to
  bottom). Also indicated is the breakdown point of the pattern at
  $\phi^*=a^*=-\log_9(a)$ (vertical line).}
\end{center}\end{figure}

\subsection{Comparison with numerical results}
\begin{figure}\begin{center}
\epsfig{figure=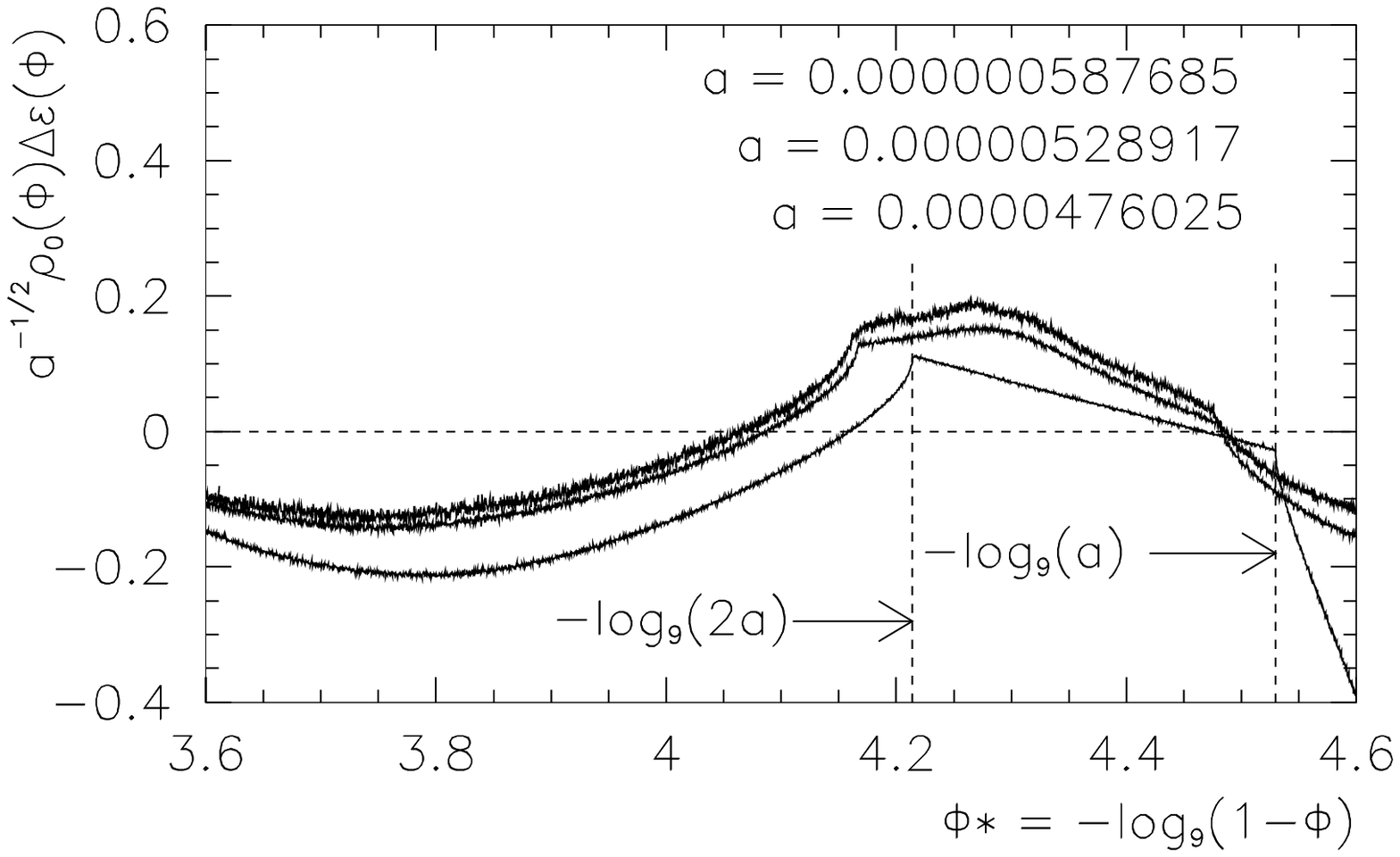, scale=0.75}
\caption{\label{ener3n}Adjusted energy difference
  $a^{-1/2}\rho_0(\phi)\Delta\epsilon(\phi)$ for the 3B-string at
  $a=0.0000476025$, $a=0.00000528917$, and $a=0.000000587685$ (from bottom to
  top).}
\end{center}\end{figure}

\begin{figure}\begin{center}
\epsfig{figure=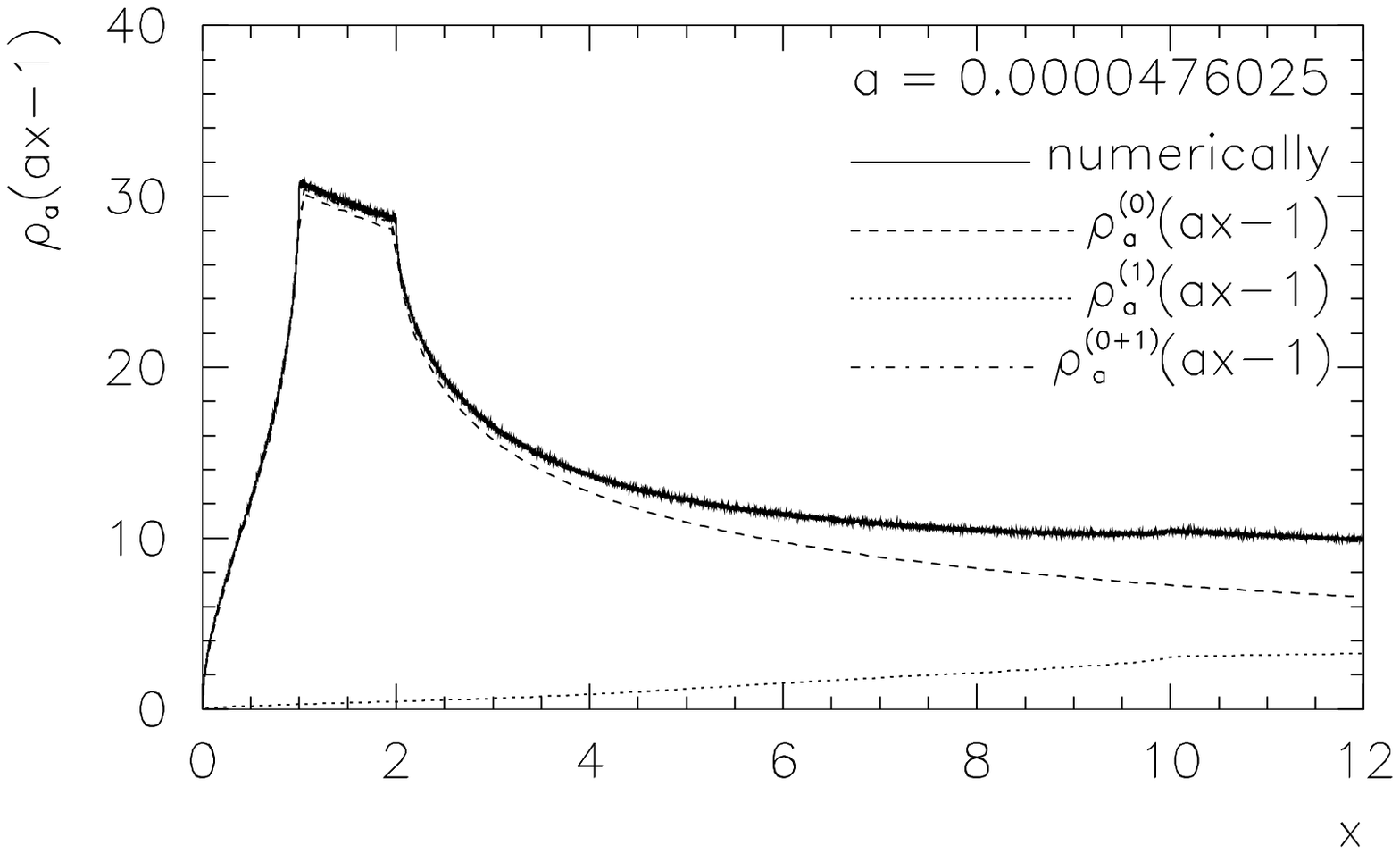, scale=0.8}
\caption{\label{iter3b}1-point density $\rho_a(1-ax)$ (solid curve) of the
  3B-string for $a=0.0000467025$ as a function of $x$, compared to the zeroth
  iterate $\rho_a^{(0)}(1-ax)$ (dashed curve), the first iterate
  $\rho_a^{(1)}(1-ax)$ (dotted curve), and the sum of both (dashed-dotted
  curve)}
\end{center}\end{figure}

As we did in Sec.~2, for numerical purposes it is useful to analyse the
adjusted energy difference $a^{-1/2}\rho_0(\phi)\Delta\epsilon(\phi)$ as a
convenient observable for which we can visualise the scaling behaviour in
$\phi^*=-\log_9(1-\phi)$ close to $\phi=+1$ (because of the symmetry of the
spectral density $\rho_a(\phi)$ in case of the 3B-string, the observation of
the region close to $\phi=-1$ gives the same result and, therefore, is
obsolete). In Fig.~\ref{ener3345} (corresponding to Fig.~\ref{ener2345} of the
2B-string) we show logarithmic plots of
$a^{-1/2}\rho_0(\phi)\Delta\epsilon(\phi)$ for the three values
$a=0.0000476025$, $a=0.00000528917$, and $a=0.000000587685$ of the coupling.
The various iterates of our convolution scheme are shown in Fig.~\ref{ener3n},
which corresponds to Fig.~\ref{ener2n}.
We observe
\begin{itemize}
\item for $p=0$ two cusps at $-\log_9(ax)=-\log_9(a)$ and $-\log_9(2a)$
\item for $p=1$ the shift of these cusps by an amount of
  $-\log_9(10/9)=-0.048\ldots$
\item for $p\geq 2$ the shift by an amount of
  $-\log_9(9/8)=-0.054\ldots$
\end{itemize}
Finally, in Fig.~\ref{iter3b} we compare the numerical result for the
invariant density $\rho_a(1-ax)$ (solid curve) with the zeroth iterate
$\rho_a^{(0)}(1-ax)$ (dashed curve), the first iterate $\rho_a^{(1)}(1-ax)$
(dotted curve), and the sum of both (dashed-dotted curve). The accuracy of the
agreement is again so high that one is not able to distinguish between the
solid and the dashed-dotted curve.

\section{Proof of the scaling behaviour in $a$}
At the end of this paper, having all the necessary results at hand, let us
finally prove that scaling in $\phi$ implies scaling in $a$. To be more
definite, we prove in the following that if $V(\phi)$ is a generic test
function, then the difference of expectation values taken with invariant
densities for small coupling $a$ and vanishing coupling $a=0$ scales with
$\sqrt{a}$ and exhibits log-periodic oscillations of period $\log N^2$:
\begin{equation}\label{VaV0}
\langle V(\phi)\rangle_a-\langle V(\phi)\rangle_0
  =\sqrt a f_V\left(\log_{N^2}(a)\right),\qquad f_V(x^*+1)=f_V(x^*).
\end{equation}
Examples of this oscillating behaviour were shown in Fig.~1.

\subsection{Scaling of $\rho_a(1-ax)$ in $a$}
To summarise, we found in section 3 that in leading order in $a$
\begin{equation}\label{rhoaxN}
\rho_a(1-ax)=\frac2{N\pi\sqrt{2a}}\sum_{p=0,1}^\infty
  \int\frac{\rho_0(\phi_+)d\phi_+\rho_0(\phi_-)d\phi_-}{N^{2p}\sqrt{x/N^{2p}
  +r_N^p(\phi_+)+r_N^p(\phi_-)}},\qquad N=2,3
\end{equation}
The sum starts at $p=0$ for $N=3$ and at $p=1$ for $N=2$. Thus we have
\begin{equation}\label{rhoaxg}
\rho_a(1-ax)=\frac1{\sqrt a}g(x)
\end{equation}
where the function $g(x)$ is independent of $a$ for small $a$.

\subsection{Scaling of $\rho_a(1-ax)$ in $\phi$}
We observed that the adjusted energy difference
$a^{-1/2}\rho_0(\phi)\Delta\epsilon(\phi)$ scales as
\begin{equation}
a^{-1/2}\rho_0(\phi)\Delta\epsilon(\phi)
  =\frac{2\pi}{N\sqrt2}f^*\left(-\log_{N^2}\pfrac{1-\phi}a\right),\qquad
  f^*(x^*-1)=f^*(x^*)
\end{equation}
close to $\phi=+1$ (the constant prefactor is chosen for later convenience).
If we write
\begin{equation}
\Delta\epsilon(\phi)=\epsilon_a(\phi)-\epsilon_0(\phi)
  =\frac1{2\pi^2\rho_0^2(\phi)}-\frac1{2\pi^2\rho_a^2(\phi)},
\end{equation}
we can expand $\rho_a(\phi)=\rho_0(\phi)\left(1-2\pi^2\rho_0^2(\phi)
  \Delta\epsilon(\phi)\right)^{-1/2}$ to obtain
\begin{eqnarray}\label{rhoarho0}
\rho_a(\phi)&\approx&\rho_0(\phi)
  \left(1+\pi^2\rho_0^2(\phi)\Delta\epsilon(\phi)\right)\ =\nonumber\\[3pt]
  &\approx&\rho_0(\phi)+\frac{2\pi^3\sqrt a}{N\sqrt 2}\rho_0^2(\phi)
  f^*\left(-\log_{N^2}\pfrac{1-\phi}a\right).
\end{eqnarray}
In order to prove the periodicity property $f^*(x^*-1)=f^*(x^*)$ we use the
fact that we can represent the invariant density $\rho_0(1-ax)$ by a series
expansion analogue to Eq.~(\ref{rhoaxN}),
\begin{eqnarray}\label{rho0axN}
\rho_0(1-ax)&=&\frac1{\pi\sqrt{1-(1-ax)^2}}\ \frac2N\sum_{p=0,1}^\infty
  \frac1{N^p}\ \approx\ \frac2{N\pi\sqrt{2a}}\sum_{p=0,1}^\infty
  \frac1{N^{2p}\sqrt{x/N^{2p}}}\ =\nonumber\\
  &=&\frac2{N\pi\sqrt{2a}}\sum_{p=0,1}^\infty\int\frac{\rho_0(\phi_+)d\phi_+
  \rho_0(\phi_-)d\phi_-}{N^{2p}\sqrt{x/N^{2p}}}.
\end{eqnarray}
In terms of $f(x)=f^*(-\log_{N^2}(x))=f^*(x^*)$ the periodicity reads
$f(N^2x)=f(x)$. The function $f(x)$ is given by
\begin{equation}
f(x)=x\sum_{p=0,1}^\infty\int\left(\frac{\rho_0(\phi_+)d\phi_+
  \rho_0(\phi_-)d\phi_-}{N^{2p}\sqrt{x/N^{2p}+r_N^p(\phi_+)+r_N^p(\phi_-)}}
  -\frac{\rho_0(\phi_+)d\phi_+\rho_0(\phi_-)d\phi_-}{N^{2p}\sqrt{x/N^{2p}}}
  \right)=\sum_{p=0,1}^\infty f_p(x).
\end{equation}
The periodicity can be shown for values $x\gg N^{2P}$ where $P$ again is
chosen large enough to guarantee the $p$-independence of the function
$r_N^p(\phi)$, $r_N^p(\phi_\pm)\approx r_N^\infty(\phi_\pm)$ for $p\ge P$. For
this range of values for $x$ we have
\begin{eqnarray}
\lefteqn{f_p(N^2x)=N^2x\int\left(\frac{\rho_0(\phi_+)d\phi_+
  \rho_0(\phi_-)d\phi_-}{N^{2p}\sqrt{N^2x/N^{2p}+r_N^\infty(\phi_+)
  +r_N^\infty(\phi_-)}}-\frac{\rho_0(\phi_+)d\phi_+
  \rho_0(\phi_-)d\phi_-}{N^{2p}\sqrt{N^2x/N^{2p}}}\right)\ =}\\
  \!\!\!&=&\!\!\!x\int\left(\frac{\rho_0(\phi_+)d\phi_+
  \rho_0(\phi_-)d\phi_-}{N^{2(p-1)}\sqrt{x/N^{2(p-1)}+r_N^\infty(\phi_+)
  +r_N^\infty(\phi_-)}}-\frac{\rho_0(\phi_+)d\phi_+
  \rho_0(\phi_-)d\phi_-}{N^{2(p-1)}\sqrt{x/N^{2(p-1)}}}\right)\ =\ f_{p-1}(x).
  \nonumber
\end{eqnarray}
On the other hand, for $p<P$ the value of $x/N^{2p}$ is large. Therefore, we
can expand the radicals,
\begin{eqnarray}
\frac1{\sqrt{x/N^{2p}+r_N^p(\phi_+)+r_N^p(\phi_-)}}-\frac1{\sqrt{x/N^{2p}}}
  &\approx&\frac1{\sqrt{x/N^{2p}}}\left(1-\frac{N^{2p}}{2x}
  \left(r_N^p(\phi_+)+r_N^p(\phi_-)\right)-1\right)\ =\nonumber\\
  &=&-\frac12\pfrac{N^{2p}}x^{3/2}\left(r_N^p(\phi_+)+r_N^p(\phi_-)\right).
\end{eqnarray}
Because
$\int\rho_0(\phi_+)d\phi_+\rho_0(\phi_-)d\phi_-(r_N^p(\phi_+)+r_N^p(\phi_-))$
remains finite, we can neglect the above difference in $f_p(x)$ for small
values of $p$. Hence, we have finally shown that
\begin{equation}\label{fN2x}
f(N^2x)=\sum_{p=0,1}^\infty f_p(N^2x)\approx\sum_{p=0,1}^\infty f_{p-1}(x)
  \approx\sum_{p=0,1}^\infty f_p(x)=f(x).
\end{equation}
We have shown that this equation holds for values $x$ with $ax\ll 1$. Note
that even though it is obvious from Figs.~\ref{ener2345} and~\ref{ener3345}
that at about $ax\approx 1$ the pattern is still periodic, we cannot prove
this by using the analytical formula~(\ref{rhoaxN}), which is a perturbative
result valid close to the upper border of the interval. Also, the
approximation we did in Eq.~(\ref{rho0axN}) no longer holds. However, if $a$
is small enough, then the region in question constitutes only a small fraction
of the entire repetitive pattern.

\subsection{Scaling of expectation values in $a$}
Let us now prove the scaling behaviour of
\begin{equation}
\langle V(\phi)\rangle_a-\langle V(\phi)\rangle_0
  =\int_{-1}^1\left(\rho_a(\phi)-\rho_0(\phi)\right)V(\phi)d\phi,
\end{equation}
where $V$ is a generic test function of the local iterates of the CML. In
order to apply the two scaling relations of the previous subsections, we split
the integral over $\phi$ into two parts by chosing a subdivision point
$\phi=1-ca$. $c$ is arbitrary but fixed (for $N=2$ e.g.\ $c=2^7=128$). We
obtain $\langle V(\phi)\rangle_a-\langle V(\phi)\rangle_0=I_u(c)+I_l(c)$ where
\begin{eqnarray}
I_u(c)&=&\int_{1-ca}^1\left(\rho_a(\phi)-\rho_0(\phi)\right)V(\phi)d\phi,\\
I_l(c)&=&\int_{-1}^{1-ca}\left(\rho_a(\phi)-\rho_0(\phi)\right)V(\phi)d\phi.
\end{eqnarray}
The fact that the integral $I_u(c)$ scales with $\sqrt{a}$ can be shown very
easily. From Eq.~(\ref{rhoaxg}) we obtain
\begin{eqnarray}
I_u(c)&=&a\int_0^c\left(\rho_a(1-ax)-\rho_0(1-ax)\right)V(1-ax)dx\nonumber\\
  &\approx&a\int_0^c\left(\frac1{\sqrt a}g(x)-\frac1{\pi\sqrt{2ax}}\right)
  V(1-ax)dx\ \approx\ \sqrt a V(1)\int_0^c\left(g(x)-\frac1{\pi\sqrt{2x}}
  \right)dx\qquad
\end{eqnarray}
where the approximations hold for $ax\ll 1$. If we replace $a$ by $\lambda a$
where $\lambda$ is (so far) an arbitrary factor and $c$ is kept constant, we
obtain
\begin{equation}
I_u(c)\to\sqrt{\lambda a}V(1)\int_0^c\left(g(x)-\frac1{\pi\sqrt{2x}}\right)dx
  =\sqrt\lambda I_u(c).
\end{equation}

For the second integral $I_l(c)$ we can assume that $\phi$ is large enough to
guarantee the preservation of the pattern given by Eq.~(\ref{fN2x}). We use
Eq.~(\ref{rhoarho0}) to obtain
\begin{equation}
I_l(c)=\sqrt a\int_{-1}^{1-ca}f^*\left(-\log_{N^2}\pfrac{1-\phi}a\right)
  \tilde V(\phi)d\phi,\qquad
  \tilde V(\phi)=\frac{2\pi V(\phi)}{N\sqrt 2(1-\phi^2)}.
\end{equation}
In replacing $a$ by $\lambda a$ we obtain
\begin{equation}
I_l(c)\to\sqrt{\lambda a}\int_{-1}^{1-\lambda ac}
  f^*\left(-\log_{N^2}\pfrac{1-\phi}a+\log_{N^2}(\lambda)\right)
  \tilde V(\phi)d\phi.
\end{equation}
In general, we do not have $I_l(c)\to\sqrt\lambda I_l(c)$ as in the case of
the integral $I_u(c)$. However, if we chose $\lambda=1/N^2$, we can use that
$f^*(x^*)$ is (nearly) periodic with period $1$ to obtain
\begin{equation}
I_l(c)\to\frac1N\sqrt a\int_{-1}^{1-ac/N^2}
  f^*\left(-\log_{N^2}\pfrac{1-\phi}a\right)\tilde V(\phi)d\phi
  =\frac1NI_l(c/N^2).
\end{equation}
If we replace the upper limit of the integral by $1-ca$, we neglect an
integration range in $\phi$ of length of order $O(a)$. Therefore, in leading
order we obtain $I_l(c)\to I_l(c)/N$. Collecting both results, we finally
obtain that under the transformation $a\to a/N^2$ we have
\begin{equation}
\langle V(\phi)\rangle_a-\langle V(\phi)\rangle_0
  \to\langle V(\phi)\rangle_{a/N^2}-\langle V(\phi)\rangle_0
  =\frac1N\Big(\langle V(\phi)\rangle_a-\langle V(\phi)\rangle_0\Big)
\end{equation}
whereas for general transformations $a\to\lambda a$ we obtain a result
different from a simple multiplication by a factor $\sqrt\lambda$. It is
obvious, then, that we have
\begin{equation}
\langle V(\phi)\rangle_a-\langle V(\phi\rangle_0
  =\sqrt af^V_N\left(\log_{N^2}(a)\right)
\end{equation}
where $f^V_N(\log_{N^2}(a))$ is a periodic function with period $1$ depending
on the test function $V(\phi)$.

\section*{Conclusion}
The treatment of nonhyperbolic CMLs exhibing chaotic behaviour is notoriously
more difficult than that of hyperbolic ones. In this paper we tackled the
problem by introducing an iterative convolution technique that provides a
perturbative expression for invariant 1-point and 2-point densities. For small
coupling parameters $a$ the invariant 1-point density $\rho_a$ of the CML is
obtained as a sum of density contributions $\rho_a^{(p)}$, where the index $p$
essentially decribes at which time in the past the nonhyperbolic region is
active in our convolution scheme. We obtained explicit perturbative
expressions for the densities of diffusively coupled Chebyshev maps of
$N$-th order. Our main examples were the cases $N=2$ and $N=3$, which
significantly differ in their symmetry properties. While in the uncoupled case
the invariant 1-point density is smooth, an arbitrarily small coupling $a$
induces a selfsimilar cascade of patterns with a variety of cusps and
non-differentiable points, which can all be understood by our perturbative
approach. We proved that arbitrary expectation values scale with the square
root of the coupling parameter and that there are log-periodic oscillations
both in the phase space and in the parameter space. Though, for reasons of
concreteness, we mainly dealt with 2nd and 3rd order Chebyshev maps, our
techniques can be applied in a similar way to other nonhyperbolic coupled
systems where the local maps exhibit fully developed chaos.

\subsection*{Acknowledgements}
This work is supported in part by the Estonian target financed project
No.~0182647s04 and by the Estonian Science Foundation under grant No.~6216.
S.~Groote acknowledges support from a grant given by the Deutsche
Forschungsgemeinschaft for staying at Mainz University as guest scientist for
a couple of months. C.~Beck's research is supported by a Springboard
Fellowship of the Engineering and Physical Sciences Research Council (EPSRC).

\end{document}